# Probing the Limit of Heat Transfer in Inorganic Crystals with Deep Learning

Jielan Li[1†], Zekun Chen[1†], Qian Wang[1†], Han Yang[1*†], Ziheng Lu[1*†], Guanzhi Li[1], Shuizhou Chen[1], Yu Zhu[1], Xixian Liu[1], Junfu Tan[1], Mingfa Tang[1], Yichi Zhou[1], Claudio Zeni[1], Andrew Fowler[1], Daniel Zügner[1], Robert Pinsler[1], Matthew Horton[1], Tian Xie[1], Tie-Yan Liu[1], Haiguang Liu[1], Tao Qin[1], Bing Lv[2], Davide Donadio[3*], Hongxia Hao[1*]

[1]Microsoft Research AI for Science.
[2]Department of Physics, University of Texas at Dallas, Richardson, TX 75080, USA.
[3]Department of Chemistry, University of California Davis, Davis, CA 95616, USA.

*Corresponding author(s). E-mail(s): hanyang@microsoft.com; zihenglu@microsoft.com; ddonadio@ucdavis.edu; hongxiahao@microsoft.com;
†These authors contributed equally to this work.

**Abstract**

Heat transfer is a fundamental property of matter. Research spanning decades has attempted to discover materials with exceptional thermal conductivity, yet the upper limit remains unknown. Using deep learning accelerated crystal structure prediction and first-principles calculation, we systematically explore the thermal conductivity landscape of inorganic crystals. We brute-force over half a million ordered crystalline structures, encompassing an extensive coverage of local energy minima in binary compounds with up to four atoms per primitive cell. We confirm diamond sets the upper bound of thermal conductivity within our search space, very likely also among all stable crystalline solids at ambient conditions. We also identify over 20 novel crystals surpassing silicon in thermal conductivity, validated by density functional theory. These include a semiconductor TaN with ultrahigh thermal conductivity($\sim$900 W m$^{-1}$ K$^{-1}$), and metallic compounds such as MnV that exhibit high lattice and electronic thermal conductivity simultaneously, a distinctive feature not observed before. These results as well as the deep learning-driven screening method, redefine the landscape of thermal transport and establish a large open-access database for future materials discovery.

# 1 Introduction

Thermal conductivity ($\kappa$) is a fundamental property of matter that characterizes its ability to conduct heat. Understanding the atomic-level mechanisms and pushing the boundaries of thermal conductivity have been central challenges in materials science.[1, 2] Discovering and engineering materials with tailored thermal conductivity, particularly those exhibiting extreme $\kappa$, is crucial for a wide range of applications, including thermal management of electronic and photonic devices, heat conductors, thermal insulators, and energy converters.[3–7]

The intriguing question of "what is the upper limit of heat transfer in matter"[1] dates back to Fourier's time and the quest for an answer has gone hand in hand with the search for extreme thermal conductors (Fig. 1).[2, 8–10] In the 1800s, metals were considered the best thermal conductors, with silver having the highest conductivity (∼430 $\mathrm{W\,m^{-1}\,K^{-1}}$, Fig. 1). In the 1920s, Peierls discovered that electrons and phonons contribute additively to $\kappa$ with the former being the dominant heat carriers in metals and the latter in insulators.[11] Diamond was identified as the material with the highest thermal conductivity (∼2000 $\mathrm{W\,m^{-1}\,K^{-1}}$ at room temperature) with phonon-dominated thermal transport.[12, 13] These mark the early progress in the quest for good thermal conductors, which was majorly driven by phenomenological models and serendipitous experiments. In the last two decades, advancements in the implementation of phonon transport theory have enabled the accurate and systematic computation of thermal conductivity from first principles.[14–17] This approach facilitated the discovery of landmark materials such as boron arsenide (BAs), with a remarkable thermal conductivity initially predicted to be 2240 $\mathrm{W\,m^{-1}\,K^{-1}}$ at room temperature[18] and later further justified with higher order phonon scattering[19, 20], and confirmed experimentally to be $1000 \pm 90\,\mathrm{W\,m^{-1}\,K^{-1}}$.[6] This discovery paved the way to the theoretically-driven search for materials with extreme thermal conductivity.[21, 22]

Despite these advances, materials with high $\kappa$ remain exceptionally rare, and the question of the ultimate limit of heat transfer in matter persists. One hypothesis is that the physical upper bound is set by diamond.[12, 23] However, it is extremely challenging to form a proof (or counterproof) to such hypothesis. A comprehensive search for high-$\kappa$ materials across the whole chemical space has been infeasible, bottlenecked by the throughput of characterizations.[24, 25] Experimentally, determining the intrinsic thermal conductivity of inorganic crystalline materials may take several years, as it requires high-quality single-crystal samples and a complex apparatus.[6, 19, 26–29] While first-principles methods offer an alternative by enabling theoretical predictions with desirable accuracy,[14, 16, 17] their computational throughput remains a limiting factor. Recent studies have reached sub-hundred throughput in computing thermal conductivities by solving Boltzmann transport equation (BTE) or running equilibrium molecular dynamics (EMD) from first principles,[30–32] but a systematic exploration of the wide space of different crystalline prototypes and chemical systems has been infeasible.

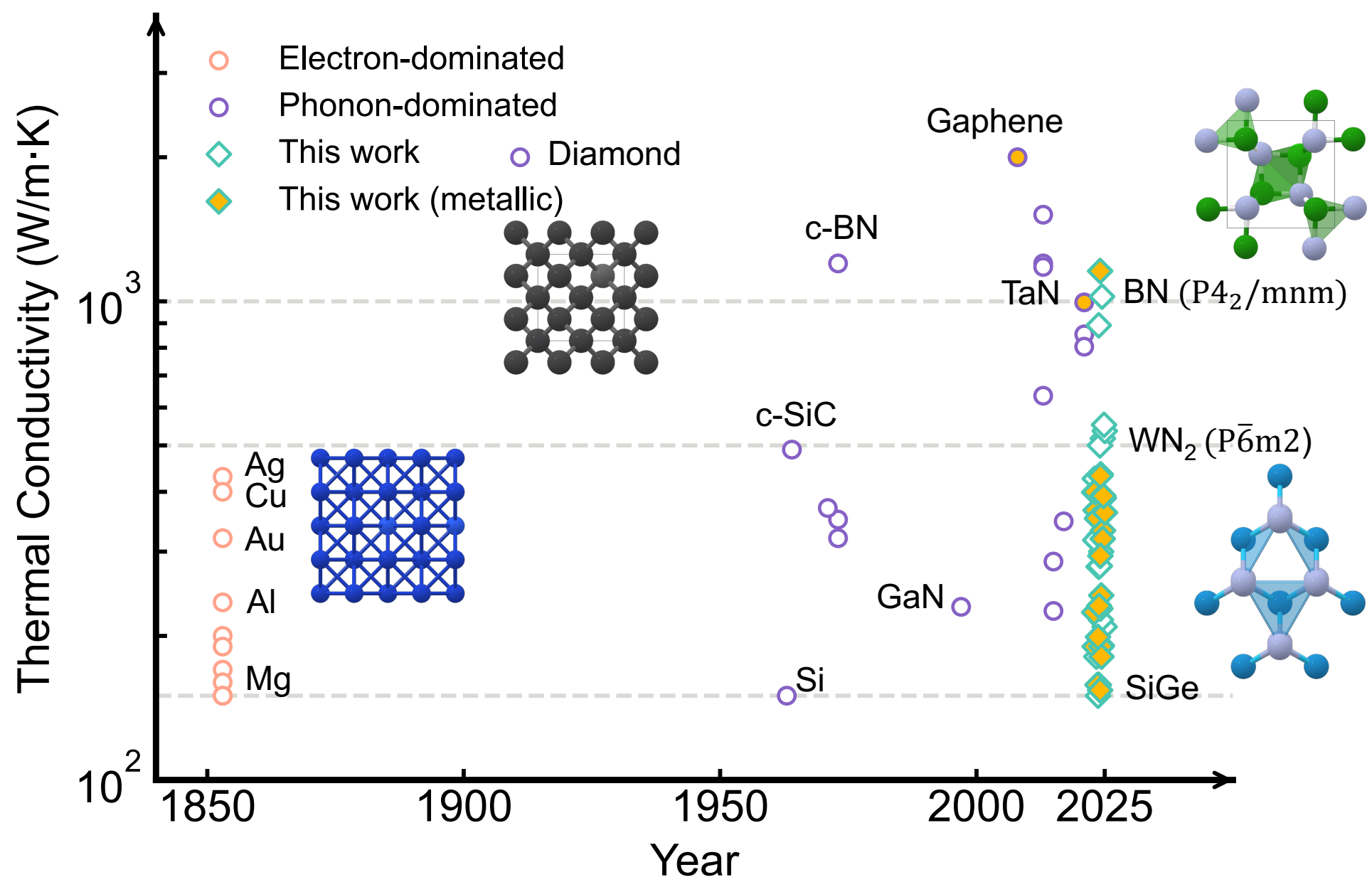

**Fig. 1**: Historical overview of high thermal conductivity materials alongside newly identified candidates from this work. Thermal conductivity values from the literature are listed in Table S5, while newly identified data are presented in Table 1, which are obtained from Boltzmann transport equation calculations based on DFT. The scattering effects considered in these calculations have been listed in Table 1. Materials represented by yellow-filled symbols indicate metallic compounds.

In this Article, we present an extensive characterization of the thermal conductivity for inorganic materials using deep learning accelerated atomistic simulations with near first-principles accuracy. In particular, we search for materials with high thermal conductivity and explore the limit of heat transfer by brute-forcing both the chemical and structural space of inorganic crystals via large-scale crystal structure prediction and high-throughput thermal conductivity calculations. A total of 642,603 structures were characterized for their vibrational properties, among which 236,574 dynamically stable crystals were studied for their thermal conductivity. The extensive search leads to the following main findings: (1) We carry out the most exhaustive search of elemental and binary ordered crystals to date, and confirm that diamond sets the upper limit of thermal conductivity of the searched structures. (2) Despite the extreme sparsity of high thermal conductors in materials space, we found over 20 novel structures (Fig. 1) with thermal conductivity exceeding that of silicon at room temperature (145 $\mathrm{W\,m^{-1}\,K^{-1}}$). (3) New binary metallic compounds with efficient phononic transport were discovered, having lattice thermal conductivity 10-fold higher than the lattice counterpart of conventional metals. In particular, the intermetallic compound MnV stands out for lattice and electronic thermal conductivity both high and of comparable magnitude, resulting in a total thermal conductivity of $243\,\mathrm{W\,m^{-1}\,K^{-1}}$, akin to aluminum.[33] This is a newly discovered distinctive thermal transport behavior. This study redefines current knowledge of thermal conductivity landscape for crystalline materials over an expansive space by providing a detailed characterization of more than 236,000 materials with near first-principles accuracy incorporating three-phonon scattering processes. The large database serves as a rich foundation for tailored design of materials to meet specific thermal management requirements for the community.

## 2 Search space for high thermal conductivity materials

We constructed a workflow to generate an extensive database of the lattice thermal transport properties of inorganic crystals, see Fig. 2a and subsection S1.1. Following Slack's seminal work,[38] we first focus on simple crystalline structures that are most likely to exhibit high thermal conductivity to challenge the diamond limit. We systematically traverse 3,240 binary chemical systems and carry out random structure searches (RSS)[39, 40] for local energy minima with up to four atoms per unit cell within these systems (Fig. S1). Given the exhaustive nature of this approach,[40] this pool effectively spans both the compositional and structural space under these constraints. A loose stability threshold of 0.2 eV/atom above the convex hull is set to avoid overlooking structures that may be kinetically stable under ambient conditions. In total, this provides a pool of 113,381 candidate structures.

In addition to this pool, we further incorporate 529,222 more complex structures with more atoms in the unit cell, more chemical species, or higher energy above the convex hull (see Fig. S1 for details). While increased structural complexity generally leads to stronger phonon scattering and reduces the probability of achieving extreme thermal conductivity, they broaden the search space, thereby increasing the likelihood of discovering novel high-$\kappa$ candidates. Adding these complex structures leads to a pool that contains 642,603 structures.

Using a deep learning atomistic model MatterSim,[41] we assess the dynamical stability of each structure within this extensive dataset under harmonic approximation. This leads to 236,574 structures that are dynamically stable for which we compute the thermal conductivity using the Boltzmann transport equation in the relaxation time approximation with three-phonon scattering. As shown in Fig. 2b, the current candidate pool is not only several orders of magnitude larger in sheer number compared with previous studies, but also warrants a much less biased sampling of both the chemical and structural space.

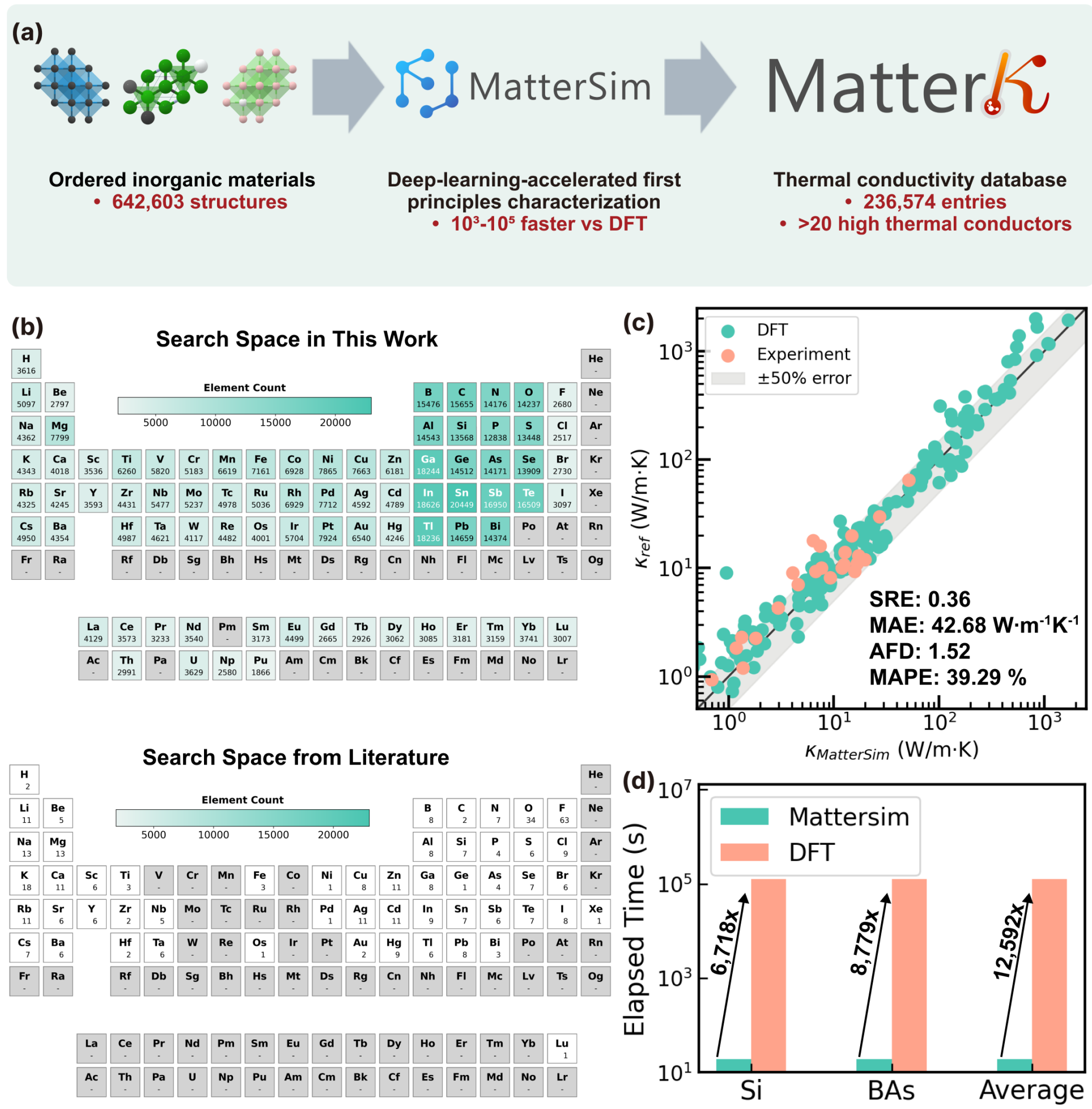

**Fig. 2**: Deep learning accelerated thermal conductivity screening. (a) Computational workflow. The initial structure database is sourced from deep learning accelerated random structure search and existing databases. Dynamically stable structures were computed for their thermal conductivity, leading to the final MatterK database; (b) Comparison of current search space between this work and previous studies[25, 32, 34–36]. (c) Parity plot comparing computed thermal conductivity and reference values. Average Factor Difference (AFD), Symmetric Relative Error (SRE)[37], Mean Absolute Error (MAE), and Mean Absolute Percentage Error (MAPE) are computed as statistical metrics for model accuracy. (d) Elapsed time computing third-order force constant by MatterSim and DFT on Si, BAs, and an average over 93 materials. One Nvidia A100-80G GPU was used for deep learning accelerated computation while two Nvidia A100-40G GPU were used for DFT computation.

## 3 Deep learning accelerated first-principles characterization

Reliable prediction of lattice thermal conductivity from atomistic simulations requires an accurate and transferable method to describe both harmonic phonon properties and higher-order anharmonic interactions.[2, 14] Density functional theory (DFT)[42, 43] or empirical potentials has been widely used for such tasks.[14, 24, 44] However, the large size of the current structure pool requires computational efficiency and high accuracy. Here, we employ MatterSim, a deep learning foundation model for atomic systems across the periodic table trained over a broad range of temperatures and pressures.[41] The model achieves state-of-the-art predictions of phonon properties of inorganic crystals[41, 45] and mitigates the phonon softening issues,[46] with high computational efficiency. During the development of this study, we notice that a few recent researches[37, 47] that also prove the feasibility of applying universal machine learning potentials to predict lattice thermal conductivity of inorganic crystals.

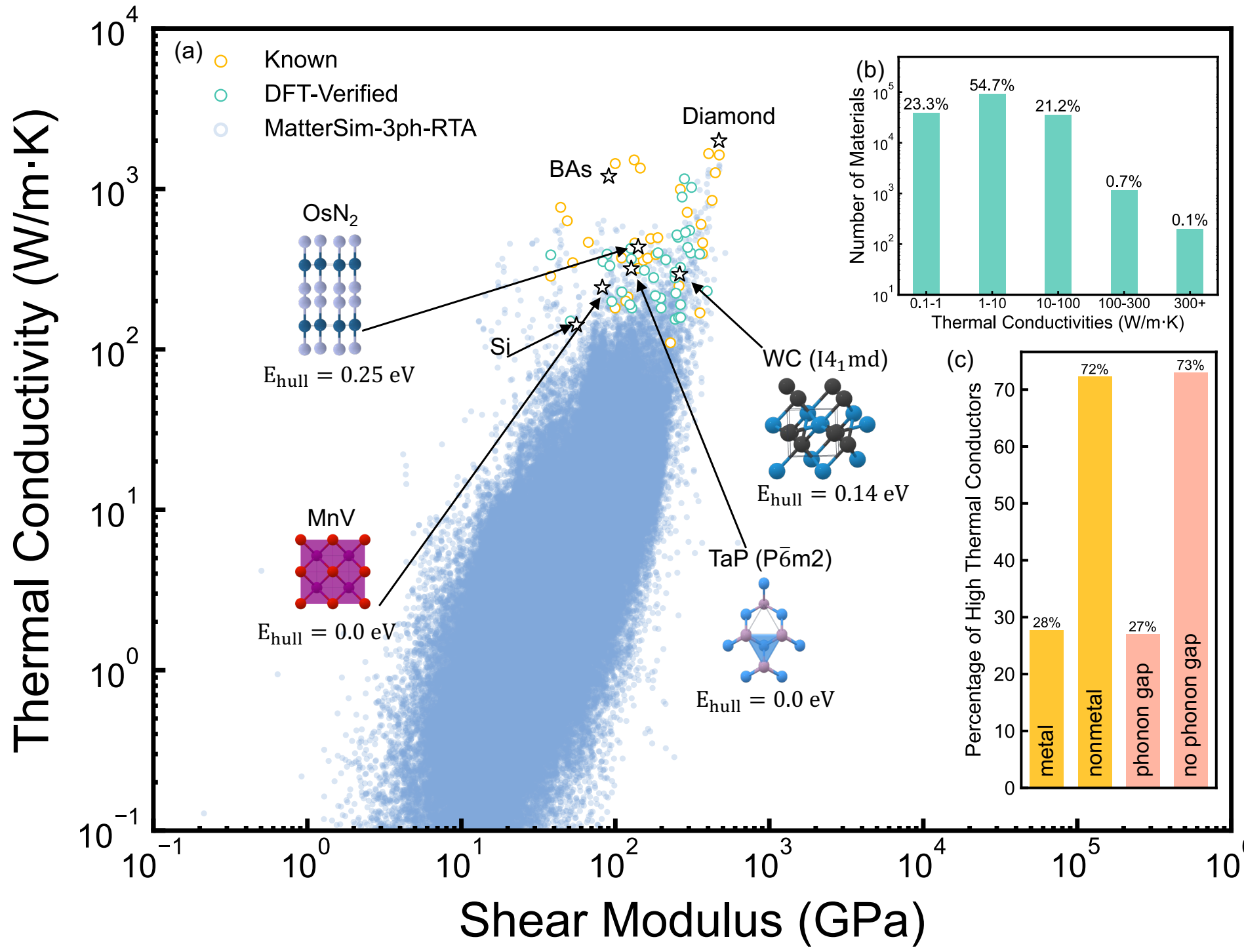

**Fig. 3**: Thermal conductivity distributions in MatterK database. (a) Lattice thermal conductivities under the three-phonon scattering process as a function of shear modulus, predicted by MatterSim for bulk crystals in the MatterK dataset (blue). Known high thermal conductors from the literature are highlighted in orange, while DFT-verified new high thermal conductor candidates are shown in purple. (b) The distribution of the lattice thermal conductivity for these structures in different ranges. (c) The classification of high thermal conductors in MatterK dataset based on whether they are metallic or non-metalllic and whether their phonon dispersion exhibit a phonon gap.

We first benchmark the model on thermal conductivity using Boltzmann transport equation with three-phonon scattering (3ph-BTE) against DFT calculations and experiments (Fig. 2c). The results demonstrate MatterSim's predictive accuracy across materials with thermal conductivities spanning from 1 to 2000 $\mathrm{W\,m^{-1}\,K^{-1}}$. This benchmark dataset, constructed through random sampling and integration of existing data, encompasses a diverse set of materials, including oxides, nitrides, and phosphides. As shown in Fig. 2c, validation against both first-principles calculations and experimental measurements reveals a strong linear correlation between MatterSim's deep learning accelerated predictions and reference values. The model achieves errors of 0.36 (symmetric relative error, SRE), $42.68\,\mathrm{W\,m^{-1}\,K^{-1}}$ (mean absolute error, MAE), 1.52 (average factor difference, AFD), and 39.29 % (mean absolute percentage error, MAPE), exhibiting high predictive accuracy,[37] especially for materials with high thermal conductivities, as detailed in Table S2. On the speed side, 3ph-BTE calculations using MatterSim (MatterSim-BTE) are at least 1,000 times faster than conventional first-principles methods. As shown in the lower panel of Fig. 2d, an average acceleration factor of up to $10^4$ times can be achieved, reducing the computational time from weeks to minutes.

Leveraging the AI-accelerated MatterSim-BTE calculator, we carried out third-order force constants and 3ph-BTE calculations for the 236,574 dynamically stable structures to determine their lattice thermal conductivity, and created MatterK, a comprehensive database of lattice thermal conductivities at the level of BTE with three-phonon scattering under the single mode relaxation time approximation (RTA). The distribution of computed thermal conductivity is plotted in Fig. 3a against the shear modulus of the materials which exhibits the strongest correlation coefficient in our Pearson's analysis (Fig. S6). Additional distributions of $\kappa$ with respect to other parameters such as symmetry are provided in Fig. S7 and Fig. S8.

# 4 The upper limit

For decades, researchers have pursued materials with exceptionally high thermal conductivity but have been unable to surpass the record set by diamond. This leads to one key hypothesis: diamond represents the fundamental physical limit for heat transfer in matter under ambient conditions. Forming a proof of this claim has been challenging — it is difficult to derive an analytical bound due to the intricacies of phonon transport and the multitude of factors governing heat conduction. We testify this hypothesis by an extensive search of possible simple crystalline structures that may stably exist as described in Section 2. We carried out 3ph-BTE calculations on all these potential structures. Such computations provide an upper boundary to $\kappa$ as they do not account for higher-order phonon scattering and other extrinsic processes. It is worthwhile to note that, a very small fraction of structures stood out from the search with $\kappa$ exceeding $500\,\mathrm{W\,m^{-1}\,K^{-1}}$, like P$\bar{6}$m2-WC[21]. Considering the deep learning error range, they could potentially break the diamond bound. A further DFT verification confirms none of these structures has thermal conductivity exceeding diamond, as shown in Fig. 3a. With these, we confirm diamond to set the limit under ambient condition within the searched structures, i.e., binary compounds with up to four atoms per unit cell.

**Table 1**: Potential high-$\kappa$ candidates verified by DFT. The structures are sourced from the Materials Project database (labeled with MP IDs) and an in-house random structure search database (labeled with RSS). The energy above the convex hull, $E_{\text{hull}}$, is reported in eV/atom, where higher values indicate greater thermodynamic instability for synthesis. The scattering process column specifies the different scattering processes considered in the DFT-BTE calculations. Determination of whether a structure is metallic or semiconductive is based on DFT calculations with with PBE functional.

| Category | Systems | ID | Space Group | $E_{\text{hull}}$ | Thermal conductivity $\kappa$ ($\mathrm{W\,m^{-1}\,K^{-1}}$) | | | |
|---|---|---|---|---|---|---|---|---|
| | | | | | xx | yy | zz | scattering process |
| | $B_6O$ | mp-1346 | R$\bar{3}$m | 0.00 | **279** | 279 | 206 | A |
| | $B_6P$ | mp-28395 | R$\bar{3}$m | 0.00 | **209** | 209 | 189 | A |
| | SiGe | mp-1219182 | F$\bar{4}$3m | 0.02 | 150 | 150 | 150 | A |
| | $WN_2$ | mp-999549 | P$\bar{6}$m2 | 0.09 | 436 | 436 | **517** | A |
| | $WN_2$ | mp-1077232 | $P6_3/mmc$ | 0.09 | 300 | 300 | **303** | A |
| | TaN | RSS | $I4_1md$ | 0.10 | **900** | 900 | 395 | B |
| | BP | RSS | R$\bar{3}$m | 0.11 | **426** | 426 | 342 | A |
| | BP | RSS | P$\bar{6}$m2 | 0.14 | **388** | 388 | 244 | A |
| | BN | mp-13151 | $P4_2/mnm$ | 0.18 | 451 | 451 | **1023** | A |
| Semiconductors and Insulators | $PtN_2$ | mp-1095618 | Pa$\bar{3}$ | 0.21 | 216 | 216 | 216 | A |
| | BN | mp-644751 | Pnma | 0.27 | 266 | **500** | 356 | A |
| | BN | mp-1077506 | Imm2 | 0.30 | 184 | 304 | **324** | A |
| | CH† | mp-1079612 | $I2_13$ | 0.32 | 536 | 536 | 536 | A |
| | $CN_2$ | mp-1009818 | I$\bar{4}$m2 | 0.73 | **399** | 399 | 216 | A |
| | $AlGaN_2$ | mp-1228894 | P3m1 | 0.01 | **181** | 181 | 144 | A |
| | AlSiCN | mp-1227998 | P3m1 | 0.04 | **317** | 317 | 275 | A |
| | $BeSiN_2$ | mp-1227309 | P3m1 | 0.17 | **280** | 280 | 102 | A |
| | $BC_2N$ | mp-1078541 | C2/m | 0.64 | 400 | **891** | 623 | A |
| | $BC_2N$ | mp-1008523 | P$\bar{4}$m2 | 1.00 | **552** | 552 | 487 | A |
| | TaP* | mp-1187244 | P$\bar{6}$m2 | 0.003 | 230 | 230 | **366** | E |
| | MnV* | mp-316 | Pm$\bar{3}$m | 0.00 | 243 | 243 | 243 | E |
| | NbB | mp-2580 | Cmcm | 0.00 | **400** | 319 | 312 | A |
| | HfS | mp-1206743 | P$\bar{6}$m2 | 0.00 | 247 | 247 | **352** | A |
| | VCr | RSS | Pm$\bar{3}$m | 0.00 | 332 | 332 | 332 | A |
| | TaP | mp-1067587 | $I4_1md$ | 0.00 | **211** | 211 | 109 | B |
| | VB | RSS | Fm$\bar{3}$m | 0.00 | 228 | 228 | 228 | A |
| | ReB | RSS | P$\bar{6}$m2 | 0.00 | 90 | 90 | **158** | D |
| | TaRe | RSS | Pm$\bar{3}$m | 0.00 | 191 | 191 | 191 | A |
| | VN | mp-1018027 | P$\bar{6}$m2 | 0.00 | 157 | **190** | 157 | A |
| | TaW | RSS | Pm$\bar{3}$m | 0.00 | 190 | 190 | 190 | A |
| | $WB_2$ | RSS | R$\bar{3}$m | 0.00 | 118 | 118 | **158** | A |
| Metallic | ReN | RSS | R3m | 0.01 | **181** | 181 | 146 | A |
| | CrC | mp-1018050 | P$\bar{6}$m2 | 0.08 | 174 | 174 | **224** | A |
| | WC | RSS | $I4_1md$ | 0.14 | **294** | 294 | 198 | D |
| | $OsN_2$ | mp-973935 | P6/mmm | 0.25 | 112 | 112 | **435** | E |
| | $BC_5$ | mp-1077125 | I$\bar{4}$m2 | 0.25 | **393** | 393 | 242 | A |
| | $BC_7$ | mp-1079046 | Pmm2 | 0.25 | **231** | 206 | 163 | A |
| | ReC | mp-1009735 | P$\bar{6}$m2 | 0.27 | 225 | 225 | **362** | A |
| | TiN | mp-998908 | F$\bar{4}$3m | 0.30 | 199 | 199 | 199 | A |
| | C | mp-1008374 | Cmmm | 0.44 | 237 | 620 | **1157** | A |
| | $MnCrV_2$ | mp-864953 | Fm$\bar{3}$m | 0.00 | 392 | 392 | 392 | A |
| | $MoWC_2$ | mp-1221393 | Pmm2 | 0.00 | **154** | 117 | 91 | D |
| | $B_2CN$ | mp-1079333 | Pmma | 0.24 | 373 | **432** | 215 | A |

∗ LDA functional is adopted in DFT calculations.
† $K_4$ phase of the carbon-hydrogen compound identified in Ref. 48.
A: 3-phonon scattering.
B: 3-phonon plus 4-phonon.
C: 3-phonon plus 4-phonon plus isotope scattering.
D: 3-phonon plus phonon-electron plus isotope scattering with electron thermal conductivity.
E: 3-phonon plus 4-phonon plus phonon-electron plus isotope scattering with electron thermal conductivity.

Beyond this space, it is difficult to further carry out extensive search for local energy minima on the potential energy surface, considering the increasing number of degrees of freedom. However, our analysis reveals a general observation that the thermal conductivity of a ternary compound tend not to exceed the highest value observed among the corresponding binary constituents (see Fig. S4 and Fig. S5). For example, while $BC_2N$ and $MoWC_2$ exhibit exceptionally high $\kappa_{\text{ph}}$ values, they remain below the maximum thermal conductivity found in their respective binary counterparts. Notably, this trend holds consistently for ternary compound with thermal conductivity exceeding $300\,\mathrm{W\,m^{-1}\,K^{-1}}$. Furthermore, when additional scattering mechanisms and electron contributions are considered, some candidates with initially high thermal conductivity, such as HfP, $MoWC_2$ and more illustrated in Table S6, undergo significant reductions in their predicted thermal conductivity values. As a result, the pool of viable ultrahigh thermal conductivity materials narrows down considerably, leaving only C, BN, and $BC_2N$ (in different phases), as confirmed by first-principles computations. Therefore, given the extensive exploration of binary systems and the constraints imposed by phonon transport physics, diamond is very likely to also set the physical limit of ambient thermal conductivity throughout the entire materials space.

## 5 High thermal conductivity materials

While it is unlikely that any bulk crystal will break the diamond limit, we identified a large pool of dynamically stable structures with potential high thermal conductivity greater than silicon. From the distribution in Fig. 3b, we observe that the vast majority of crystalline materials exhibit lattice thermal conductivities in the range of $1\,\mathrm{W\,m^{-1}\,K^{-1}}$ to $10\,\mathrm{W\,m^{-1}\,K^{-1}}$, with fewer than 1% exceeding $100\,\mathrm{W\,m^{-1}\,K^{-1}}$ Fig. 3b. This rarity underscores why, despite centuries of search, compounds with exceptionally high thermal conductivity have remained elusive. Nevertheless, under three-phonon approximation, we identified 968 structures with lattice thermal conductivities exceeding that of silicon, representing the largest dataset of high thermal conductivity crystalline materials compiled to date. In particular, 112 of them show a thermal conductivity over $500\,\mathrm{W\,m^{-1}\,K^{-1}}$. Many of these structures belong to compositional and structural ranges previously unknown. For example, a set of high thermal conductivity candidates was identified with previously unreported chemical compositions with distinct structure features, including $OsN_2$, $MoWC_2$, MnV, ReB, $I2_13$-CH[48], $Be_2CoNi$, and $WN_2$. We also identified a unique family of materials, exemplified by $OsN_2$, that share a one-dimensional (1D)-chain-like structure with P6 symmetry. These materials, which include $OsC_2$, $ReN_2$, RuN, NbF, CoN, TiF, TcF, and IrN, exhibit ultrahigh thermal conductivity along the z-direction. Unfortunately, they are typically characterized by high formation energies, suggesting substantial thermodynamic instability.

From the extensive set of 968 deep learning-prescreened high-$\kappa$ structures, we have validated ~20 most promising newly identified thermal conductors through first-principles three phonon-BTE calculations - a number comparable to the total count of historically known high thermal conductivity materials (Table S5). Among these, a noteworthy case is a tetragonal phase of TaN uncovered from our random structure search. DFT calculations at the PBE level reveal this phase to be a semiconductor with an estimated band gap of approximately 0.6 eV. It exhibits exceptionally high thermal conductivity of 900, 900,and 395 $\mathrm{W\,m^{-1}\,K^{-1}}$ along the three lattice directions when accounting for four-phonon scattering processes. Notably, we noticed a contemporaneous report of this phase during the preparation of this manuscript,[49] further supporting the significance of this discovery. More broadly, a substantial fraction of these materials exhibits phonon gaps, similar to BAs, while a significant proportion are metallic (Fig. 3c). These findings highlight the necessity of incorporating higher-order phonon interactions and electron-phonon coupling in thermal transport models. Given the high computational costs to assess four-phonon and electron-phonon scattering by first principles, we performed full calculation for only a subset of these materials (Table 1), acknowledging that $\kappa$ of some currently identified high thermal conductivity candidates may be overestimated once all scattering mechanisms are considered.

## 6 Metallic compounds with high lattice thermal conductivity

The thermal conductivity of crystalline materials sums up the contributions of two energy carriers: electrons and quantized lattice vibrations, i.e. phonons. Conventional metals such as copper ($400\,\mathrm{W\,m^{-1}\,K^{-1}}$), aluminum ($237\,\mathrm{W\,m^{-1}\,K^{-1}}$), and iron ($80\,\mathrm{W\,m^{-1}\,K^{-1}}$) exhibit high electronic thermal conductivity ($\kappa_{\mathrm{el}}$) but very low phononic conductivity ($\kappa_{\mathrm{ph}}$). In contrast, a small subset of metallic and semi-metallic compounds have been recently found to have anomalously high $\kappa_{\mathrm{ph}}$ but relatively low $\kappa_{\mathrm{el}}$, including $\theta$-TaN,[22] NbN,[50] P$\bar{6}$m2-WC,[21] I$4_1$md-WC (this work), and P$\bar{6}$m2-TaP (this work). However, materials with significant contributions from both electron and phonon transport remain missing, to the best of our knowledge.

We identify two metallic compounds MnV and $OsN_2$ that simultaneously exhibit both exceptionally high $\kappa_{\mathrm{el}}$ and $\kappa_{\mathrm{ph}}$ (Fig. 4a). Notably, MnV has $\kappa_{\mathrm{el}}$ $\sim$$100\,\mathrm{W\,m^{-1}\,K^{-1}}$, and a lattice thermal conductivity of $120\,\mathrm{W\,m^{-1}\,K^{-1}}$ computed from first principles, a relatively high value for a metallic compound consisting of two heavy atoms. The unique characteristics of this material arise from a combination of several factors, as explained below. MnV crystallizes in a *bcc* structure, that features 3-fold degeneracy of both acoustic and optical phonons at the high-symmetry $R$ point, with only a small frequency difference between the two branches (Fig. 4b). Such patterns in phonon dispersion effectively reduce three-phonon scattering processes by limiting the available phase space for energy and momentum conservation, thereby enhancing phonon transport efficiency (Fig. 4d-e). A similar mechanism due to degenerate acoustic phonons has also been reported in the *bcc* phases of metals W, Mo and Cr in Ref. 51. As a further comparison, we analyzed TiFe and ScCo with the same crystal structure as MnV but have larger frequency difference between acoustic and optical branches near $R$ point, see Fig. 4b and Fig. 4d. These three materials exhibit similar group velocities across the entire Brillouin zone; however, their lifetimes differ by several orders of magnitude near the $R$ point. As a result, the thermal conductivity contributions near the $R$ point decrease significantly from MnV to

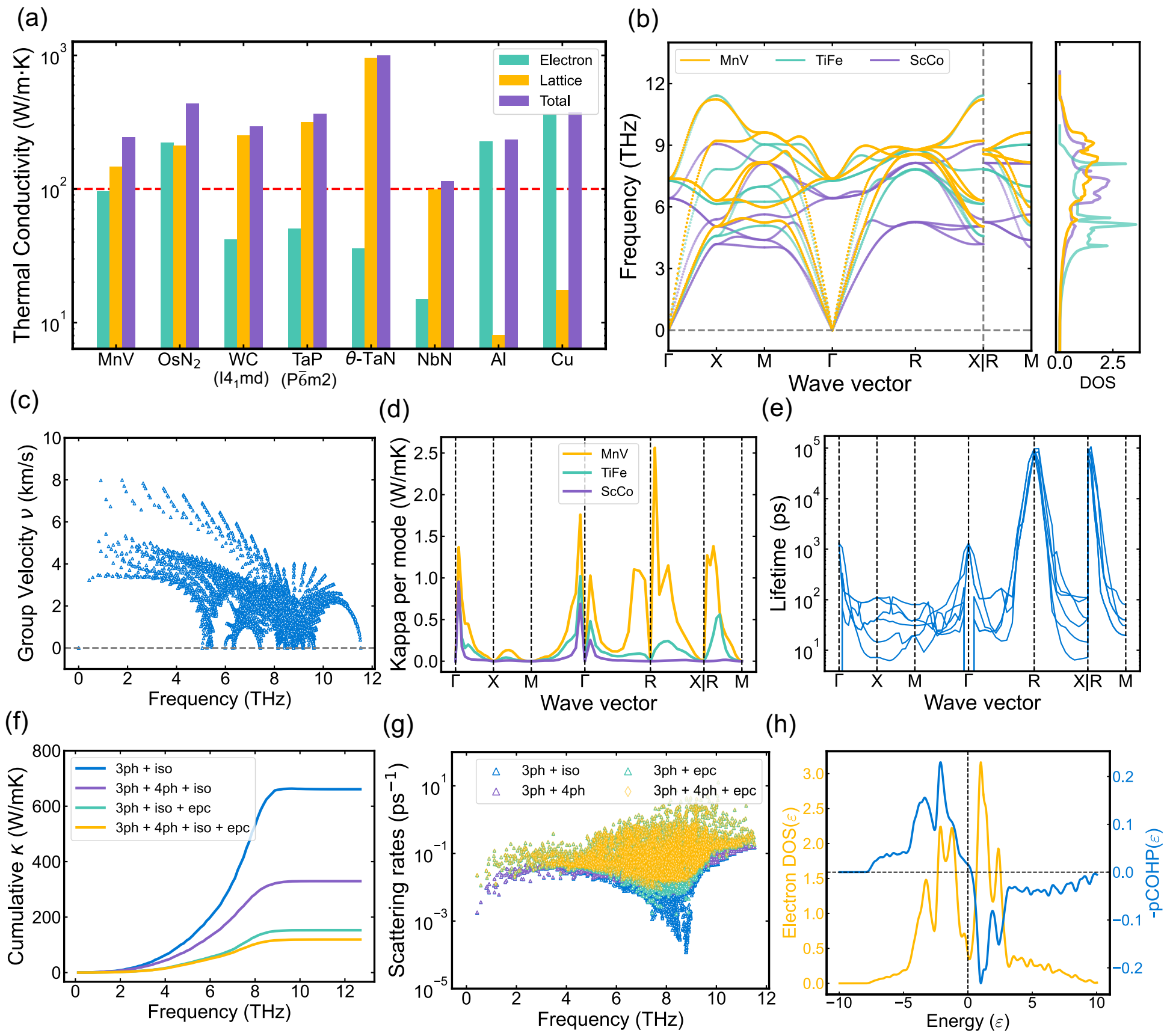


**Fig. 4**: (a) The electron part, lattice part and total of thermal conductivities for metallic systems with high thermal conductivity. (b) Phonon dispersion and density of states for MnV, TiFe and ScCo. (c) Group velocity for MnV. (d) $\kappa$ per mode along high symmetry points path for MnV, TiFe and ScCo. (e) Lifetime of different phonon bands along high symmetry points path for MnV.(f) Cumulative thermal conductivities for MnV considering different scattering mechanism ("3ph", "4ph", "iso" and "epc" represent three-phonon process, four-phonon process, isotopes effect and electron-phonon coupling effect, respectively). (g) Scattering rates for MnV. (h) Electronic density of states (DOS) and projected crystal orbital Hamilton population (pCOHP) for MnV.

TiFe and become negligible for ScCo (Fig. S12). This trend aligns with the increasing acoustic-optical frequency difference observed near the $R$ point.

Electron-phonon scattering processes are known to hamper phonon transport in conventional metals. Interestingly, despite the metallic nature, the electron-phonon scattering in MnV only reduces its phonon thermal conductivity $\kappa_{\mathrm{ph}}$ by a factor of four as shown in Fig. 4f, contrasting a reduction of over a factor of ten observed in a few recently reported metallic compounds (e.g. TiN, HfN, NbC) that were also predicted to have high thermal conductivity with three-phonon processes.[52, 53] Electronic structure analysis reveals that the limited electron-phonon scattering originates from a sharp minimum of the electronic density of states (DOS) at the Fermi energy, see Fig. 4h and Fig. S9c-d. This electronic distribution leads to strong Mn–V bonds, as evidenced by the projected crystal orbital

Hamilton population (pCOHP)[54, 55] which displays a rather distinct separation of bonding and anti-bonding states and a sharp drop of pCOHP values at the Fermi energy. This electronic configuration yields stiff bonds with high vibrational frequencies and maintains the metallic characteristic of the material, yet with moderate reduction of lattice thermal conductivity due to electron-phonon scattering. These characteristics lead to the unusually even balance between electronic and phononic heat transport. Such a feature may have significant implications for interconnect in semiconductor devices as it would reduce the thermal boundary resistance between metallic contacts and semiconducting layers.

## 7 Discussions

We conducted a large-scale search for high thermal conductivity materials using deep learning accelerated atomistic simulations to understand the limits of heat transport in solids and identify new thermal conductors. By systematically exploring the chemical and structural space of ordered inorganic crystals, we characterized 642,603 structures and computed lattice thermal conductivity for 236,574 dynamically stable ones. We confirm diamond to have the highest thermal conductivity among all searched structures and, most likely, in the overall space of bulk crystals, at ambient conditions. While not breaking the diamond limit, the search uncovered over 20 new high-$\kappa$ candidates surpassing silicon with many in previously unexplored chemical and structural spaces, providing valuable references for advanced thermal management applications. Notably, we identified a class of metallic materials exhibiting both high electron- and phonon-mediated thermal conductivity, an exotic property previously not observed to the best of our knowledge. These findings not only expand the known landscape of thermal materials but also challenge conventional assumptions regarding phonon transport in metallic systems. Furthermore, this work offers the largest dataset to date for materials informatics and thermal management design.

Despite these findings, several limitations should be noted. Our search comprises two parts: an unbiased extensive sampling of simple crystals and an *ad hoc* selection of more complex structures. As a result, the upper bound for thermal conductivity is conclusively determined only for searched structures, although we expect this conclusion to hold more generally across the entire inorganic bulk crystalline space. The deep learning model, trained on large-scale density functional theory data, reproduces the overall energy landscape and achieves state-of-the-art accuracy in lattice thermal conductivity computations but still incurs an average error on the order of tens of $\mathrm{W\,m^{-1}\,K^{-1}}$, potentially causing missed structures in the $150 \sim 300\,\mathrm{W\,m^{-1}\,K^{-1}}$ range. Additionally, our screening, based on three-phonon scattering, may overestimate thermal conductivity, leading to discrepancies after incorporating all scattering processes. Lastly, while DFT using the PBE or LDA functional

generally performs well, it has known limitations. Band gaps are systematically underestimated,[56–59] leading to the prediction that some small-gap semiconductors are metallic. The classification of materials in this paper into metals or non-metals in this paper are subject to this error, and as such the classification could be wrong in a small number of cases. Additionally, some of the structures are not on the convex hull, particularly those arising from the extensive random structure search, meaning that they are not predicted to be thermodynamically stable under the assumptions of the convex hull construction. It is likely that some structures will not be synthesizable or very difficult to synthesize, including those that depend on very specific ordering of species, the less energetically favorable polymorphs of well-studied material systems, and the unusual structures that are predicted to only be stable under certain conditions, such as the theoretically predicted $OsN_2$ phase.[60] This is because dynamical stability is a necessary but not sufficient condition for synthesizability, and many dynamically stable structures cannot be synthesized. Despite this, the MatterK database provides a comprehensive collection of material candidates that can be used to guide further experimental efforts.

This work also highlights new opportunities in thermal conduction research enabled by deep learning. This work addresses the upper bound of heat conduction under ambient conditions, leveraging the insight that phonons are less prone to scattering in simple crystals, allowing a most exhaustive search to date with deep learning accelerated simulations. In contrast, determining the lower bound remains more complex,[61–66] as structurally intricate materials are likely to dominate, necessitating advanced search strategies such as generative models and others.[67, 68] Furthermore, the large-scale dataset generated in this study provides a valuable resource for future investigations. Systematic data mining may reveal new physical correlations and facilitate the discovery of materials optimized for diverse thermal management applications.

# 8 Methods

## 8.1 Chemical Search Space

The search pool is generated in-house through an AI-accelerated random structure search (RSS) and supplemented with data from the Materials Project (MP) database[69]. In this study, the terms “binary” and “ternary” systems are defined as follows. For binary systems, an exhaustive crystal structure search is conducted for compositions with up to four atoms per unit cell, considering elements with atomic number lower than 95, while excluding Group 18 elements, the highly radioactive element 61, 84–89 and 91. Given the computational cost and complexity of ternary systems, an exhaustive search across all possible ternary compositions is impractical. Since the lattice thermal conductivity of metals is normally low due to electron-phonon scattering and a large number of atoms

in the unit cell leads to higher impact of three-phonon scattering, for ternary compounds we limit the RSS search to systems made of main-group elements 13-16 and fewer than 8 atoms per cell. This choice allows us to optimize search efficiency while capturing relevant chemical diversity.The RSS settings are the same as Ref.[41]. When an $E_{\text{hull}}$ threshold of 0.2 eV/atom was later applied for identifying stable materials, the convex hull was constructed using the MP database v2022.10.28, while the $E_{\text{hull}}$ values for the RSS structures were computed using MatterSim potentials.

### 8.2 Thermal conductivity workflow

All calculations were performed using the near first-principles foundational deep learning model, MatterSim. The computation of thermal conductivity was systematically carried out in four stages.

First, structural relaxation and filtering were conducted using the Fast Inertial Relaxation Engine (FIRE) algorithm to ensure that atomic forces remained below 0.001 eV/Å. Redundant and non-3D structures were subsequently excluded, with dimensionality identified using the Rank Determination Algorithm (RDA)[70] implemented in ASE.[71]

Following structural relaxation, phonon dispersion relations were computed using Phonopy.[72] Any system exhibiting imaginary frequencies above $10\,\text{cm}^{-1}$ was discarded, as the presence of imaginary frequencies at 0 K indicates structural instability. Such instabilities typically require temperature-dependent renormalization of both second- and third-order force constants, often addressed through effective potential fitting or a self-consistent approach.[73–75] However, incorporating temperature-dependent effects into high-throughput screening would significantly increase computational cost and is therefore beyond the scope of this work.

For systems found to be dynamically stable, thermal conductivity calculations were performed using 3ph-BTE, as detailed in subsection 8.3.

### 8.3 ML-accelerated force constants for BTE

The second-order interatomic force constants and phonon spectra are computed using Phonopy,[72] interfaced with the machine-learning potential MatterSim. The magnitude of finite displacements is set to 0.03 Å.

The third-order force constants are computed using Phono3py,[72] maintaining the same supercell size and displacement magnitude as in the second-order calculations. No cutoff is imposed on the atomic distance when supercells are created for third-order force constants. The thermal conductivity is then solved using Phono3py within the Relaxation Time Approximation (RTA). To ensure convergence, the q-point mesh for Brillouin zone integration is set to six times the supercell size.

## 8.4 Predicting Mechanical Properties Using MatterSim

Elastic constants describe the stiffness of a material in response to small deformations (i.e., within the elastic regime). For isotropic materials, common moduli include Young's modulus $E$, shear modulus $G$, bulk modulus $B$, and Poisson's ratio $\nu$. For anisotropic crystals, the elastic behavior is described by a full $6 \times 6$ symmetric matrix (expressed in Voigt notation) of independent elastic constants $C_{ij}$. In the regime of small strains, Hooke's Law can be written as

$$\sigma_i = \sum_j C_{ij}\, \epsilon_j,$$

where $\sigma_i$ are the stress components, $\epsilon_j$ are the strain components, and $C_{ij}$ are the elastic constants. Furthermore, the change in energy due to strain can be expressed by a second-order Taylor expansion:

$$\Delta E = \frac{1}{2} V \sum_{i,j} C_{ij}\, \epsilon_i\, \epsilon_j,$$

which indicates that the elastic constants are essentially the second derivatives of the total energy with respect to strain. Since MatterSim provides the total energy, forces, and stress of a system, both the stress-strain method and the energy-strain method can, in principle, be used to compute the mechanical properties. However, the energy-strain method offers several advantages. For example, it is generally less sensitive to numerical noise in the calculated stresses and often yields more stable and convergent results, particularly when dealing with subtle energy differences. Moreover, the energy-strain method allows for a direct determination of the elastic constants as the second derivatives of the total energy with respect to strain. The energy-strain approach has been implemented in the VASPKIT software package[76]. To leverage this capability, we have developed an interface between MatterSim and VASPKIT, enabling efficient calculation of mechanical properties using the energy-strain method.

## 8.5 DFT validation

### 8.5.1 DFT 3ph

All force constants needed to solve the phonon BTE are calculated by projector-augmented plane wave method[77] as implemented in VASP,[78, 79] and the atomic displacement amplitude for finite displacement are set to 0.01 Å. The Generalized Gradient Approximation (GGA) Perdew-Burke-Ernzerhof (PBE) exchange-correlation functional[80] is used throughout the work unless stated otherwise. A 0.18 $\text{Å}^{-1}$ $k$ points spacing is used to sample the first Brillouin Zone, and the plane-wave energy cut-off is set to 520 eV. We first perform structural optimization to ensure that the forces

on the atoms in the system are less than $10^{-3}$ eV/Å. Subsequently, we use the Phonopy[72] and ShengBTE[81] packages to calculate the second-order and third-order force constants, respectively. To ensure that the second-order and third-order force constants converge as much as possible, we require that the lattice constants along all three directions of the supercell are greater than 10 Å, and the number of atoms in the supercell is no less than 100. In the calculation of third-order force constants, we selected the cut-off to the sixth nearest neighbor atom to balance computational efficiency and accuracy. For the calculation of Boltzmann transport equation, we test multiple $q$ mesh points to obtain convergent lattice thermal conductivity values.

### 8.5.2 DFT 4ph

For the calculation of fourth-order force constants, the FourPhonon package[82] is used to generate displaced supercell and to compute the fourth-phonon interactions. We used a smaller supercell compared to the third-order force constant calculation, with one fewer unit cell in each direction, and the cut-off was set to the 2nd nearest neighbor atom.

### 8.5.3 DFT electron-phonon coupling

All electron-phonon related DFT calculation are performed using Quantum Espresso (QE) [83][84] package. The pseudopotential are used in pseudo-dojo[85] with Optimized norm-conserving Vanderbilt pseudopotentials[86] and GGA exchange-correlation functional. The $k$ points to sample first Brillouin Zone are same as VASP used, and the plane-wave energy cut-off is set to 80 Ry. For lattice thermal conductivity limited by electron scattering, we employ the EPW [87][88] package along with QE to calculate phonon lifetime. And the modified ShengBTE package were used to include the phonon-electron scattering to calculate $\kappa_{ph}$. For electron thermal conductivity, we employ the PERTURBO[89] package based on maximally localized Wannier function[90] to calculate $\kappa_{el}$ in dense $k, q$ mesh to ensure convergence. For the Wannier interpolation used in EPW and PERTURBO, we carefully select the projector and energy window for each system to ensure that the Wannier function spread of each system is less than a lattice constant, and the interpolation values on the energy band near the Fermi surface and phonon spectrum are the same as those calculated by DFT.

### 8.5.4 Calculations of DOS and pCOHP

We calculated the electronic density state (DOS) and the projected crystal orbital Hamilton population (pCOHP) as shown in Fig. S9 using Lobster[54] package (version 5.11). Specifically, following a self-consistent field (SCF) calculation using VASP with a $k$-mesh density of $0.04 \times \pi$ $\text{Å}^{-1}$[76], the resulting output was processed with LOBSTER: The lobster input file was configured with an energy window from -10 to 10 eV relative to the Fermi energy (using **COHPStartEnergy** -10 and

**COHPENDEnergy** 10) and defined the local basis by including $s$, $p$ and $d$ orbitals using the **includeOrbitals** command. To probe the atomic-resolved and orbital-resolved interactions between Mn and V, we generated the bonding interactions for all Mn-V pairs within a distance range of 1.0 to 3.0 Å, employing the **cohpGernator** command. The eDOS and pCOHP values are computed following this protocol.

# 9 Code Availability

The MatterSim model used in this work is already publicly available at: MatterSim repository.

# 10 Acknowledgements

We thank Chris Bishop, Bin Shao, Ryota Tomioka, Jia Zhang, Karin Strauss for their invaluable support. We appreciate David Broido and David Cahill for insightful discussions. We acknowledge Bichlien Nguyen, Jake Smith for constructive feedbacks for the work; Deniz Gunceler and Maik Riechert for their support with our computational infrastructure; Jingyun Bai for improving the quality of the figures; Shoko Ueda, Peggy Dai and Roberto Sordillo for their instrumental management of the project. We are grateful to Kenji Takeda, Shruti Rajurkar, and others for their support in establishing external collaborations and overall study coordination. Additionally, HH thanks Wu Li, Yiyang Sun and Haibo Xiao for instructive discussions. HY thanks Cunzhi Zhang, Giulia Galli, Yi Xia, for their thoughtful suggestions during private discussions. ZC thanks Bohan Li, Wang-Yeuk Kong, Dylan Folkner, Kam-Tung Chan and Frank Cerasoli for fruitful discussions.

# 11 Author Contributions

JL, HY, ZL and HH conceived the study, JL, ZC, QW, HY, ZL, GL, SC, YZ, XL, JT, MT,YZ implemented the methods and workflows, JL, ZC, QW, HY, ZL, GL, SC, YZ, MT, CZ, AF performed experiments, CZ, AF, DZ, RP, MH, TX helped with the implementation of the methods and all authors participate discussions and wrote this manuscript. HH led the research.

# Declarations

While drafting this manuscript, we utilized the language model GPT-4 to facilitate sentence-level composition and to ensure clarity and coherence.

# Supplementary Information: Probing the Limit of Heat Transfer in Inorganic Crystals with Deep Learning

Jielan Li[1†], Zekun Chen[1†], Qian Wang[1†], Han Yang[1*†], Ziheng Lu[1*†], Guanzhi Li[1], Shuizhou Chen[1], Yu Zhu[1], Xixian Liu[1], Junfu Tan[1], Mingfa Tang[1], Yichi Zhou[1], Claudio Zeni[1], Andrew Fowler[1], Daniel Zügner[1], Robert Pinsler[1], Matthew Horton[1], Tian Xie[1], Tie-Yan Liu[1], Haiguang Liu[1], Tao Qin[1], Bing Lv[2], Davide Donadio[3*], Hongxia Hao[1*]

[1]Microsoft Research AI for Science.
[2]Department of Physics, University of Texas at Dallas, Richardson, TX 75080, USA.
[3]Department of Chemistry, University of California Davis, Davis, CA 95616, USA.

*Corresponding author(s). E-mail(s): hanyang@microsoft.com; zihenglu@microsoft.com; ddonadio@ucdavis.edu; hongxiahao@microsoft.com;
†These authors contributed equally to this work.

In addition to this supporting information document, some other supplementary files will be attached: a) Unconfirmed high thermal conductors b) MatterK database c) All ternary compounds and the corresponding highest binary compounds. d) DFT verification setting.

## Contents

# S1 Search space

## S1.1 Search space definition

Fig. S1 illustrates the construction details of the chemical search space and their contributions to the final Matter$\kappa$ database where high thermal conductors most likely reside. The in-house designed target search space resulted in > 6M unique crystal structure entries, which is too large to be fully screened using MatterSim-BTE. Here we chose to conduct an extensive screening using MatterSim-BTE for a subset of the materials, i.e., binary and ternary systems with up to 4 atoms in the unit cell and $E_{\mathrm{hull}} \leq 0.2$ eV/atom, aiming to effectly traverse the local energy minima and explore the heat limit for stable materials under ambient conditions. For the other subsets, some ad-hoc screening rules/constraints are applied to select the more relevant material entries followed by MatterSim-BTE screening. The few subsets are: (1) binary and ternary materials composed of either 6 or 8 atoms in the unit cell with $E_{\mathrm{hull}} \leq 0.2$ eV/atom, with/out filtered by bulk modulus higher than 100GPa; (2) binary materials composed of up-to-8 atoms per unit cell with $E_{\mathrm{hull}} > 0.2$ eV/atom, selecting the lowest-energy entry for each unique composition and unique symmetry; (3) ternary materials composed of up-to-8 atoms in the unit cell with $E_{\mathrm{hull}} > 0.2$ eV/atom, selecting the ten lowest-energy entries for each unique composition; (4) BCN system for all $E_{\mathrm{hull}}$ ranges. (5)Materials Project database with up-to-16 atoms per unit cell.

## S1.2 Thermal conductivity distribution from Materials Project

To validate the criteria used in defining the search space as above, we screened the Materials Project[1] database with minimal bias (up-to-16 atoms per unit cell) using MatterSim screening, where 3ph-BTE computation is conducted for dynamically stable materials after geometry optimization. The results indicate that 95% of materials with thermal conductivity exceeding that of silicon ($145\,\mathrm{W\,m^{-1}\,K^{-1}}$) belong to up-to-ternary chemical systems with no more than eight atoms per unit cell (Fig. S2). Notably, no materials meeting this thermal conductivity threshold were found in systems with five or seven atoms per unit cell, and the candidates that have more than four atoms in the unit cell

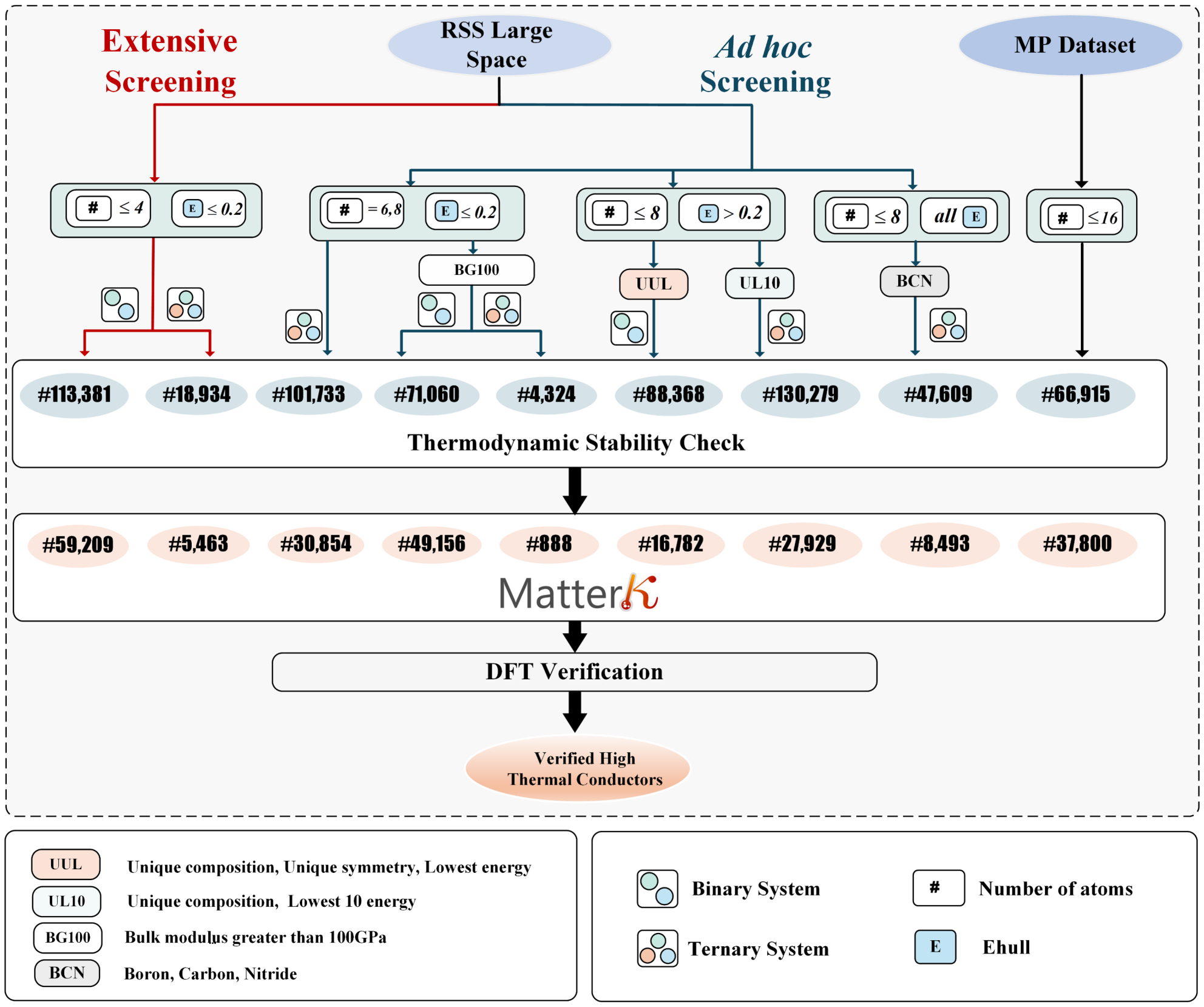


**Fig. S1**: Detailed illustration of the chemical search space generated from random structure search (RSS) and Materials Project (MP) and selection rules designed to identify regions where high thermal conductors are most likely to be found. The final selected structures undergo thermodynamic stability check followed by MatterSim-BTE calculations, forming the MatterK database. The number in blue circles indicate the structure counts from each category that will go through the thermodynamic stability check through harmonic lattice dynamics. The number in orange circles indicate the structure counts from each category that go through MatterSim-BTE calculator and contribute to the MatterK database. The selection rules are listed in the bottom left.

but possess thermal conductivity exceeding $500\,\mathrm{W\,m^{-1}\,K^{-1}}$ all belong to the B-C-N chemical system. These findings validate our RSS space design strategy, which focuses on binary and ternary systems with up to eight atoms per unit cell while excluding five- and seven-atom configurations, and MatterSim-BTE screening focus more on simple structures with at most four atoms per unit cell. It's worth noting the high thermal conductivity material identified with 16 atoms per unit cell is a diamond-like carbon structure. We anticipate that similar high thermal conductivity materials may exist in systems with larger unit cells, such as 32 or 64 atoms; however, exploring these systems is beyond the scope of this work.

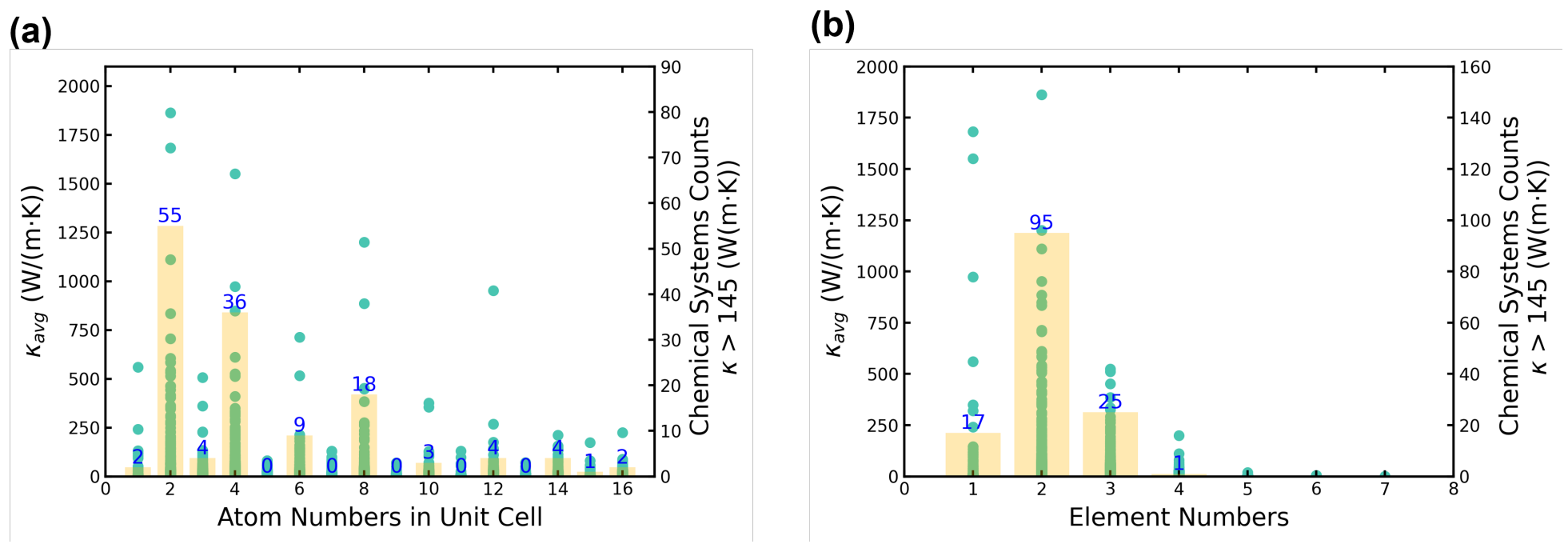


**Fig. S2**: Lattice thermal conductivity $\kappa$ distribution analysis for Materials Project database screened using MatterSim-BTE. (a) $\kappa$ vs number of atoms in unit cell. (b) $\kappa$ vs number of elements in unit cell. The dots represent the distribution of all structures within each category, while the bars and numbers indicate the high thermal conductivity candidates ($> 145\,\mathrm{W\,m^{-1}\,K^{-1}}$) in that category.

# S2 Benchmarks

## S2.1 Three-Phonon Boltzmann Transport Equation

Here we proceed with a thorough and systematic benchmark on the $\kappa$ predictions when coupling MatterSim with the three-phonon Boltzmann transport equation (3ph-BTE). BTE captures nuclear quantum effects and is much more computationally affordable compared to equilibium molecular dynamics (EMD). Thus, BTE has been the method of choice for materials with high thermal conductivity[2–4]. In this work, we compared a set of 120 materials, of which their $\kappa$ were either probed experimentally or validated with first-principle based BTE calculations[5–31]. The raw data for the benchmark plot in Fig. 2c is shown in Table S3.

To evaluate $\kappa$ predictions across different orders of magnitude, we report a few error metrics here.

(1) Average factor difference (AFD). In the context of thermal transport, AFD can be defined as follow:

$$10^x; x = \frac{1}{N}\sum_i^N |\log_{10}(\kappa_{\ell,i}) - \log_{10}(\kappa_{ref,i})|, \tag{1}$$

where $N$ represents the number of benchmarking systems, and $\kappa$ denotes the MatterSim predictions, which are derived from BTE. Several studies have adopted the use of AFD in high-throughput screening of $\kappa$ using end-to-end models, where the ground truth labels are sourced from experimental measurements[32–34].

(2) Symmetric Relative Error(SRE)

$$\mathrm{SRE}[\kappa] = \frac{1}{N}\sum_i^N 2\frac{|\kappa_{\ell,\mathrm{i}} - \kappa_{\mathrm{ref,i}}|}{\kappa_{\ell,i} + \kappa_{ref,i}}. \tag{2}$$

**Table S1**: Performance of `M3GNet-MP`, `CHGNet`, `Mattersim`, `MACE-MPA`, `OMAT24`, `SevenNet`, `GRACE-2L-OMA` and `MACE-OMAT`, on prediction of $\kappa$ on the in-house $\kappa$ dataset listed in Table S3.

| MAE of | M3GNet-MP | CHGNet | Mattersim | MACE-MPA-0 | OMAT24 | SevenNet | GRACE-2L-OMA | MACE-OMAT |
|---|---|---|---|---|---|---|---|---|
| **SRE**[$\kappa$] | 1.33 | 1.81 | 0.36 | 0.35 | 1.89 | 0.69 | **0.20** | 0.22 |
| **MAE**[$\kappa$]($\mathrm{W\,m^{-1}\,K^{-1}}$) | 105.69 | 118.93 | 42.68 | 132.55 | 1.3e5 | 76.89 | **41.74** | 93.94 |
| **AFD**[$\kappa$] | 10.78 | 161.37 | 1.52 | **1.48** | 2.7e4 | 2.56 | 2.98 | 2.41 |
| **MAPE**[$\kappa$](%) | 77.37 | 94.43 | **39.29** | 50.30 | 849498.29 | 50.41 | 48.76 | 42.53 |

**Table S2**: Performance of `M3GNet-MP`, `CHGNet`, `Mattersim`, `MACE-MPA`, `OMAT24`, `SevenNet`, `GRACE-2L-OMA` and `MACE-OMAT`, on prediction of $\kappa$ on the **high** thermal conductivity materials (in-house $\kappa$ dataset with $\kappa_{ref}$ higher than $100\,\mathrm{W\,m^{-1}\,K^{-1}}$) listed in Table S3.

| MAE of | M3GNet-MP | CHGNet | Mattersim | MACE-MPA-0 | OMAT24 | SevenNet | GRACE-2L-OMA | MACE-OMAT |
|---|---|---|---|---|---|---|---|---|
| **SRE**[$\kappa$] | 1.59 | 1.95 | 0.37 | 0.52 | 1.93 | 0.88 | 0.38 | **0.33** |
| **MAE**[$\kappa$]($\mathrm{W\,m^{-1}\,K^{-1}}$) | 457.85 | 510.03 | **187.86** | 624.19 | 1.9e4 | 343.20 | 191.17 | 443.28 |
| **AFD**[$\kappa$] | 24.03 | 604.81 | **1.57** | 1.89 | 1124434.33 | 3.29 | 8.28 | 5.93 |
| **MAPE**[$\kappa$](%) | 86.06 | 98.51 | **27.95** | 71.82 | 6605.99 | 59.66 | 28.16 | 41.03 |

(3) mean absolute error(MAE) and mean absolute percentage error(MAPE)

$$\mathrm{MAE}[\kappa] = \frac{1}{N}\sum_{i}^{N} |\kappa_{\ell,i} - \kappa_{ref,i}| \tag{3}$$

$$\mathrm{MPAE}[\kappa] = \frac{1}{N}\sum_{i}^{N} \frac{|\kappa_{\ell,i} - \kappa_{ref,i}|}{\kappa_{ref,i}} \tag{4}$$

## S2.2 Elastic Moduli Predictions from MatterSim

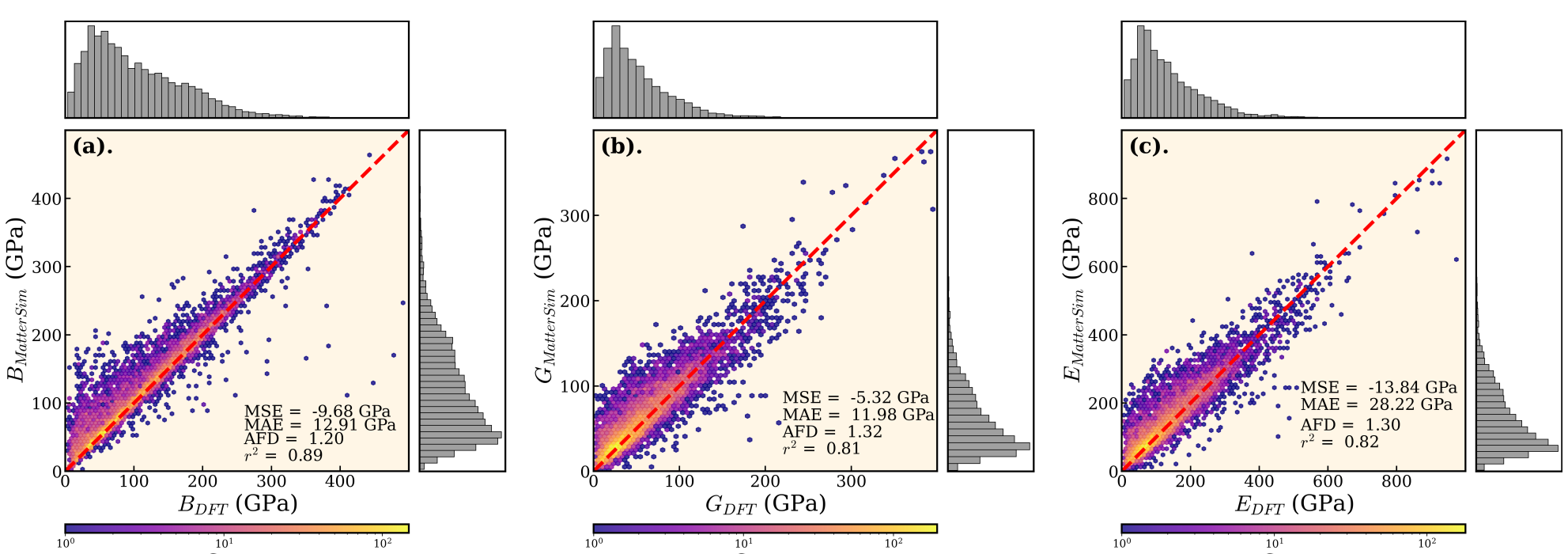

**Fig. S3**: Prediction performance from MatterSim on various elastic properties: (a) Bulk modulus ($B$), (b) shear modulus ($G$), and (C) Young's modulus ($E$) in comparison to DFT calculations. DFT data is adopted from Material project[35, 36].

**Table S3**: Summary of BTE benchmark. Rank of materials is determined based on $\kappa^{BTE}_{MatterSim}$ predicted under RTA. DFT-based BTE reference calculations performed in-house is indicated as **This work**. "h-" notation implies the hexagonal shape of the unit cell. $\kappa$ values shown in the table are $W\ m^{-1}\ K^{-1}$. RS, ZB, and WZ represent Rock-salt, Zinc-blende, and Wurtzite, respectively.

| Rank | Materials | mp-id | $\kappa^{BTE}_{MatterSim}$ | $\kappa_{ref}$ | $\kappa_{ref}$ Type | Source |
|---|---|---|---|---|---|---|
| 1 | C | mp-66 | 1683 | 1957 | Theory | [37] |
| 2 | BN | mp-1639 | 1109 | 1170 | Theory | [21] |
| 3 | BAs | mp-10044 | 850.1 | 1680.162 | Theory | **This work** |
| 4 | h-BN | mp-2653 | 846 | 927 | Theory | [21] |
| 5 | GeC | mp-1002164 | 548.4 | 1099.3 | Theory | [20] |
| 6 | $BC_2N$ | mp-30148 | 525 | 853 | Theory | [37] |
| 7 | BSb | mp-997618 | 472.7 | 537.2 | Theory | **This work** |
| 8 | SiC | mp-8062 | 465 | 415 | Theory | [21] |
| 9 | $BC_2N$ | mp-629458 | 451.6 | 804 | Theory | [37] |
| 10 | h-SiC | mp-7140 | 377 | 375 | Theory | [21] |
| 11 | BeO | mp-1778 | 352 | 358 | Theory | [21] |
| 12 | BP | mp-1479 | 275.4 | 414.7 | Theory | [27] |
| 13 | h-BeO | mp-2542 | 273.5 | 292 | Theory | [21] |
| 14 | AlN | mp-1700 | 232.6 | 212 | Theory | [21] |
| 15 | h-GaN | mp-804 | 206.7 | 225.5 | Theory | [21] |
| 16 | GaN | mp-830 | 199.6 | 231 | Theory | [21] |
| 17 | GaP | mp-2490 | 178.3 | 118.1 | Theory | [27] |
| 18 | BeTe | mp-252 | 173.7 | 253.5 | Theory | **This work** |
| 19 | BeSe | mp-1541 | 168 | 226.7 | Theory | **This work** |
| 20 | Si | mp-149 | 145.1 | 145 | Theory | [38] |
| 21 | BeS | mp-442 | 139 | 157 | Theory | [21] |
| 22 | $C_3N_4$ | mp-2852 | 133.6 | 229 | Theory | [37] |
| 23 | pseudo$-C_3N_4$ | mp-571653 | 128.6 | 262 | Theory | [37] |
| 24 | AlSb | mp-2624 | 127 | 86.6 | Theory | [27] |
| 25 | h-GaP | mp-8882 | 103.1 | 97.4 | Theory | [21] |
| 26 | $Be_2C$ | mp-1569 | 94.6 | 112.0 | Theory | [37] |
| 27 | InN | mp-20411 | 77.8 | 95.4 | Theory | [21] |
| 28 | InP | mp-20351 | 74.4 | 89.1 | Theory | [27] |

Continued on next page

**Table S3**: (continued)

| Rank | Materials | mp-id | $\kappa^{BTE}_{MatterSim}$ | $\kappa_{ref}$ | $\kappa_{ref}$ Type | Source |
|---|---|---|---|---|---|---|
| 29 | AlP | mp-1550 | 71.9 | 83 | Theory | [21] |
| 30 | ZnO | mp-1986 | 66.4 | 58.2 | Theory | [21] |
| 31 | h-InP | mp-966800 | 61.9 | 67.2 | Theory | [21] |
| 32 | h-AlP | mp-8880 | 60.2 | 73.1 | Theory | [21] |
| 33 | h-ZnO | mp-2133 | 52.8 | 52.4 | Theory | [21] |
| 34 | MgO | mp-1265 | 51.5 | 55.4 | Theory | [27] |
| 35 | Ge | mp-32 | 51.5 | 65 | Experiment | [15] |
| 36 | ZnS | mp-10695 | 50.1 | 41.8 | Theory | [27] |
| 37 | ZnSe | mp-1190 | 31.2 | 19.0 | Theory | [27] |
| 38 | GaAs | mp-2534 | 30.6 | 31.8 | Theory | [21] |
| 39 | MgSe | mp-13031 | 30.4 | 20 | Theory | [31] |
| 40 | NiO | mp-19009 | 27.2 | 30 | Experiment | [5] |
| 41 | MgTe | mp-13033 | 26.5 | 15.7 | Theory | [21] |
| 42 | GaSb | mp-1156 | 26.1 | 32.5 | Theory | [27] |
| 43 | InAs | mp-20305 | 25.9 | 27.9 | Theory | [27] |
| 44 | CaS | mp-1672 | 24.2 | 33.7 | Theory | [29] |
| 45 | CaO | mp-2605 | 23.0 | 23.2 | Theory | [27] |
| 46 | h-GaAs | mp-8883 | 22.6 | 27.3 | Theory | [21] |
| 47 | h-ZnSe | mp-380 | 20.7 | 15.2 | Theory | [21] |
| 48 | TiCoSb | mp-5967 | 19.8 | 12 | Experiment | [14] |
| 49 | MgS | mp-1315 | 19.5 | 22.59 | Theory | [29] |
| 50 | LiSiB | mp-1100392 | 18.5 | 23.45 | Theory | [30] |
| 51 | h-InAs | mp-1007652 | 18.5 | 17.7 | Theory | [21] |
| 52 | NaF | mp-682 | 18.35 | 23.5 | Theory | [27] |
| 53 | VFeSb | mp-567636 | 17.1 | 13.0 | Experiment | [13] |
| 54 | $Li_2O$ | mp-1960 | 16.7 | 11.0 | Experiment | [9] |
| 55 | h-CdS | mp-672 | 16.35 | 19 | Theory | [21] |
| 56 | TiNiSn | mp-924130 | 15.8 | 9.3 | Experiment | [12] |
| 57 | InSb | mp-20012 | 14.8 | 20.0 | Experiment | [15] |
| 58 | h-ZnTe | mp-8884 | 14.6 | 15.0 | Theory | [21] |

Continued on next page

**Table S3**: (continued)

| Rank | Materials | mp-id | $\kappa^{BTE}_{MatterSim}$ | $\kappa_{ref}$ | $\kappa_{ref}$ Type | Source |
|---|---|---|---|---|---|---|
| 59 | CdSe | mp-2691 | 14.5 | 10.3 | Theory | [21] |
| 60 | SrSe | mp-2758 | 14.3 | 19.36 | Theory | [29] |
| 61 | $ThO_2$ | mp-643 | 12.7 | 14.0 | Experiment | [16] |
| 62 | CaSe | mp-1415 | 12.4 | 15.56 | Theory | [29] |
| 63 | $CeO_2$ | mp-20194 | 12.4 | 10.8 | Experiment | [39] |
| 64 | CdS | mp-2469 | 12.4 | 21.3 | Theory | [21] |
| 65 | $SrSiO_3$ | mp-1017439 | 11.9 | 10.1 | Experiment | [19] |
| 66 | NaH | mp-23870 | 11.9 | 15.41 | Theory | [24] |
| 67 | LiH | mp-23703 | 11.7 | 14.9 | Theory | [27] |
| 68 | LiF | mp-1138 | 11.5 | 13.3 | Theory | [27] |
| 69 | h-CdSe | mp-1070 | 11.4 | 9.22 | Theory | [21] |
| 70 | HfSnPd | mp-11869 | 10.75 | 9.32 | Theory | [30] |
| 71 | $Mg_2Si$ | mp-1367 | 9.25 | 8.2 | Experiment | [6, 7] |
| 72 | h-CuH | mp-24093 | 9.23 | 7.32 | Theory | [21] |
| 73 | SrO | mp-2472 | 9.05 | 10.0 | Theory | [27] |
| 74 | CaTe | mp-1519 | 8.91 | 10.13 | Theory | [29] |
| 75 | MgSe | mp-10760 | 8.63 | 10.98 | Theory | [29] |
| 76 | $CaF_2$ | mp-2741 | 8.61 | 9.76 | Theory | [27] |
| 77 | CdTe | mp-406 | 8.16 | 7.55 | Theory | [27] |
| 78 | $KMgF_3$ | mp-3448 | 8.04 | 8.25 | Theory | [23] |
| 79 | SrS | mp-1087 | 7.86 | 12.81 | Theory | [29] |
| 80 | $CoSb_3$ | mp-1317 | 7.65 | 10.0 | Experiment | [11] |
| 81 | HgTe | mp-2730 | 7.61 | 4.62 | Theory | **This work** |
| 82 | h-CuI | mp-569346 | 7.54 | 6.51 | Theory | [21] |
| 83 | $Sb_3Ir$ | mp-1239 | 7.34 | 16.0 | Experiment | [18] |
| 84 | h-CdTe | mp-12779 | 7.09 | 5.35 | Theory | [21] |
| 85 | NbCoSn | mp-1094088 | 6.93 | 12.23 | Theory | [30] |
| 86 | $Mg_2Ge$ | mp-408 | 6.73 | 9.3 | Experiment | [6, 7] |
| 87 | $Mg_2Pb$ | mp-20724 | 6.36 | 18.0 | Experiment | [8] |
| 88 | NaCl | mp-22862 | 6.29 | 8.45 | Theory | [27] |

Continued on next page

**Table S3**: (continued)

| Rank | Materials | mp-id | $\kappa_{MatterSim}^{BTE}$ | $\kappa_{ref}$ | $\kappa_{ref}$ Type | Source |
|---|---|---|---|---|---|---|
| 89 | SrTe | mp-1958 | 6.14 | 10.16 | Theory | [29] |
| 90 | MgTe | mp-1008786 | 5.9 | 4.45 | Theory | [29] |
| 91 | ZnO | mp-2229 | 5.79 | 6.7 | Theory | **This work** |
| 92 | KCl | mp-23193 | 5.36 | 7.75 | Theory | [27] |
| 93 | $RbMgF_3$ | mp-8402 | 4.64 | 4.54 | Theory | [23] |
| 94 | KF | mp-463 | 4.61 | 6.5 | Theory | [27] |
| 95 | GeTe | mp-938 | 4.56 | 2.34 | Theory | [40] |
| 96 | $Mg_2Sn$ | mp-2343 | 4.553 | 7.1 | Experiment | [6, 7] |
| 97 | BaS | mp-1500 | 4.44 | 3.07 | Theory | **This work** |
| 98 | $SrF_2$ | mp-981 | 4.06 | 9.035 | Experiment | [17] |
| 99 | $CsCaF_3$ | mp-7104 | 3.05 | 3.03 | Theory | [23] |
| 100 | HgS | mp-1123 | 3.03 | 4.9 | Theory | [27] |
| 101 | $CdF_2$ | mp-241 | 2.96 | 4.3 | Experiment | [17] |
| 102 | $LiCl_2$ | mp-22905 | 2.74 | 3.5 | Theory | [22] |
| 103 | BaO | mp-1342 | 2.23 | 3.6 | Theory | [27] |
| 104 | AgI | mp-22925 | 2.19 | 2.02 | Theory | [21] |
| 105 | $CsCdF_3$ | mp-8399 | 1.87 | 1.73 | Theory | [23] |
| 106 | RbF | mp-11718 | 1.8 | 2.27 | Experiment | [15] |
| 107 | $BaLiF_3$ | mp-10250 | 1.77 | 2.21 | Theory | [23] |
| 108 | PbSe | mp-2201 | 1.47 | 1.55 | Theory | [27] |
| 109 | NaBr | mp-22916 | 1.43 | 3.25 | Theory | [27] |
| 110 | $Ca_5Al_2Sb_6$ | mp-8439 | 1.36 | 1.2 | Experiment | [28] |
| 111 | $SrCl_2$ | mp-23209 | 1.32 | 2.3 | Experiment | [41] |
| 112 | LiBr | mp-23259 | 1.18 | 1.83 | Experiment | [15] |
| 113 | PbTe | mp-19717 | 1.075 | 2.15 | Theory | [27] |
| 114 | NaI | mp-23268 | 1.03 | 2.1 | Theory | [27] |
| 115 | *SnSe | mp-691 | 0.78 | 0.8 | Theory | [26] |
| 116 | CsBr | mp-22906 | 0.69 | 0.94 | Experiment | [42] |
| 117 | $CsHgF_3$ | mp-561947 | 0.671 | 1.0 | Theory | [23] |
| 118 | TlBr | mp-568560 | 0.43 | 0.58 | Experiment | [28] |

Continued on next page

**Table S3**: (continued)

| Rank | Materials | mp-id | $\kappa^{BTE}_{MatterSim}$ | $\kappa_{ref}$ | $\kappa_{ref}$ Type | Source |
|---|---|---|---|---|---|---|
| 119 | $Sn_2S_3$ | mp-1509 | 0.064 | 0.06 | Theory | [25] |
| 120 | RS-AgBr | mp-23231 | 0.24 | 0.25 | Theory | [43] |
| 121 | WZ-AgI | mp-580941 | 1.30 | 1.42 | Theory | [43] |
| 122 | ZB-AgI | mp-22925 | 1.90 | 2.22 | Theory | [43] |
| 123 | RS-AgCl | mp-22922 | 0.23 | 0.29 | Theory | [43] |
| 124 | WZ-AlAs | mp-8881 | 129.36 | 71.97 | Theory | [43] |
| 125 | ZB-AlAs | mp-2172 | 163.28 | 87.41 | Theory | [43] |
| 126 | RS-BaO | mp-1342 | 2.07 | 2.69 | Theory | [43] |
| 127 | WZ-AlN | mp-661 | 225.37 | 229.73 | Theory | [43] |
| 128 | ZB-AlN | mp-1700 | 228.76 | 222.00 | Theory | [43] |
| 129 | RS-BaS | mp-1500 | 4.60 | 5.26 | Theory | [43] |
| 130 | WZ-AlSb | mp-1018100 | 100.18 | 61.42 | Theory | [43] |
| 131 | ZB-AlP | mp-1550 | 70.14 | 86.19 | Theory | [43] |
| 132 | RS-BaSe | mp-1253 | 0.95 | 9.04 | Theory | [43] |
| 133 | WZ-BAs | mp-984718 | 578.34 | 1395.87 | Theory | [43] |
| 134 | ZB-AlSb | mp-2624 | 124.50 | 100.12 | Theory | [43] |
| 135 | RS-BaTe | mp-1000 | 0.16 | 9.25 | Theory | [43] |
| 136 | WZ-BP | mp-1008559 | 266.76 | 422.05 | Theory | [43] |
| 137 | ZB-BAs | mp-10044 | 828.65 | 2004.66 | Theory | [43] |
| 138 | RS-CaO | mp-2605 | 23.85 | 24.00 | Theory | [43] |
| 139 | WZ-BeS | no-mp-4 | 90.82 | 130.84 | Theory | [43] |
| 140 | ZB-BN | mp-1639 | 1075.16 | 1188.44 | Theory | [43] |
| 141 | RS-CaS | mp-1672 | 25.73 | 29.73 | Theory | [43] |
| 142 | WZ-BeSe | no-mp-3 | 102.45 | 314.99 | Theory | [43] |
| 143 | ZB-BP | mp-1479 | 274.93 | 429.80 | Theory | [43] |
| 144 | RS-CaSe | mp-1415 | 12.96 | 16.39 | Theory | [43] |
| 145 | WZ-BeTe | mp-1183441 | 184.37 | 180.06 | Theory | [43] |
| 146 | ZB-BeO | mp-1778 | 340.14 | 363.85 | Theory | [43] |
| 147 | RS-CaTe | mp-1519 | 8.13 | 8.42 | Theory | [43] |
| 148 | WZ-CuBr | no-mp-1 | 1.09 | 1.82 | Theory | [43] |

Continued on next page

**Table S3**: (continued)

| Rank | Materials | mp-id | $\kappa_{MatterSim}^{BTE}$ | $\kappa_{ref}$ | $\kappa_{ref}$ Type | Source |
|---|---|---|---|---|---|---|
| 149 | ZB-BeS | mp-422 | 129.61 | 161.72 | Theory | [43] |
| 150 | RS-CdO | mp-1132 | 5.27 | 6.17 | Theory | [43] |
| 151 | WZ-CuCl | mp-1184046 | 0.44 | 0.83 | Theory | [43] |
| 152 | ZB-BeSe | mp-1541 | 175.88 | 383.34 | Theory | [43] |
| 153 | RS-CsF | mp-1784 | 0.85 | 1.34 | Theory | [43] |
| 154 | WZ-GaSb | mp-1018059 | 17.71 | 23.74 | Theory | [43] |
| 155 | ZB-BeTe | mp-252 | 187.05 | 282.23 | Theory | [43] |
| 156 | RS-KBr | mp-23251 | 0.89 | 2.45 | Theory | [43] |
| 157 | WZ-InN | mp-22205 | 72.72 | 110.78 | Theory | [43] |
| 158 | ZB-CdS | mp-2469 | 16.78 | 23.21 | Theory | [43] |
| 159 | RS-KCl | mp-23193 | 6.21 | 6.24 | Theory | [43] |
| 160 | WZ-InSb | mp-1007661 | 12.73 | 9.75 | Theory | [43] |
| 161 | ZB-CdSe | mp-2691 | 13.98 | 12.37 | Theory | [43] |
| 162 | RS-KF | mp-463 | 4.65 | 6.45 | Theory | [43] |
| 163 | WZ-MgTe | mp-1039 | 19.19 | 12.08 | Theory | [43] |
| 164 | ZB-CdTe | mp-406 | 7.93 | 6.95 | Theory | [43] |
| 165 | RS-KH | mp-24084 | 7.12 | 8.80 | Theory | [43] |
| 166 | WZ-ZnS | mp-561286 | 33.09 | 39.50 | Theory | [43] |
| 167 | ZB-CuBr | mp-22914 | 1.41 | 2.95 | Theory | [43] |
| 168 | RS-KI | mp-22898 | 0.08 | 1.25 | Theory | [43] |
| 169 | ZB-CuCl | mp-22905 | 0.53 | 1.44 | Theory | [43] |
| 170 | RS-LiBr | mp-23259 | 1.16 | 1.66 | Theory | [43] |
| 171 | ZB-CuH | no-mp-2 | 21.99 | 12.20 | Theory | [43] |
| 172 | RS-LiCl | mp-22905 | 3.19 | 3.50 | Theory | [43] |
| 173 | ZB-CuI | mp-22895 | 9.13 | 6.94 | Theory | [43] |
| 174 | RS-LiF | mp-1138 | 13.08 | 14.04 | Theory | [43] |
| 175 | ZB-GaAs | mp-2534 | 29.23 | 32.53 | Theory | [43] |
| 176 | RS-LiH | mp-23703 | 11.44 | 27.27 | Theory | [43] |
| 177 | ZB-GaN | mp-830 | 185.03 | 230.32 | Theory | [43] |
| 178 | RS-LiI | mp-22899 | 0.46 | 1.23 | Theory | [43] |

Continued on next page

**Table S3**: (continued)

| Rank | Materials | mp-id | $\kappa^{BTE}_{MatterSim}$ | $\kappa_{ref}$ | $\kappa_{ref}$ Type | Source |
|---|---|---|---|---|---|---|
| 179 | ZB-GaP | mp-2490 | 153.30 | 96.23 | Theory | [43] |
| 180 | RS-MgO | mp-1265 | 53.29 | 53.35 | Theory | [43] |
| 181 | ZB-GaSb | mp-1156 | 24.16 | 32.65 | Theory | [43] |
| 182 | RS-NaBr | mp-22916 | 1.36 | 2.43 | Theory | [43] |
| 183 | ZB-InAs | mp-20305 | 24.24 | 20.55 | Theory | [43] |
| 184 | RS-NaCl | mp-22862 | 6.23 | 7.13 | Theory | [43] |
| 185 | ZB-InN | mp-20411 | 76.08 | 99.05 | Theory | [43] |
| 186 | RS-NaF | mp-682 | 19.17 | 22.18 | Theory | [43] |
| 187 | ZB-InP | mp-20351 | 73.85 | 74.10 | Theory | [43] |
| 188 | RS-NaH | mp-23870 | 11.41 | 14.86 | Theory | [43] |
| 189 | ZB-InSb | mp-20012 | 14.45 | 12.66 | Theory | [43] |
| 190 | RS-NaI | mp-23268 | 0.98 | 1.27 | Theory | [43] |
| 191 | ZB-MgTe | mp-13033 | 28.39 | 16.39 | Theory | [43] |
| 192 | RS-PbS | mp-21276 | 1.72 | 1.21 | Theory | [43] |
| 193 | ZB-SiC | mp-8062 | 416.03 | 447.12 | Theory | [43] |
| 194 | RS-PbSe | mp-2201 | 1.08 | 0.73 | Theory | [43] |
| 195 | ZB-ZnO | mp-1986 | 68.36 | 63.99 | Theory | [43] |
| 196 | RS-PbTe | mp-19717 | 1.12 | 0.87 | Theory | [43] |
| 197 | ZB-ZnS | mp-10695 | 39.02 | 41.02 | Theory | [43] |
| 198 | RS-RbBr | mp-22867 | 0.33 | 3.12 | Theory | [43] |
| 199 | ZB-ZnSe | mp-1190 | 26.96 | 17.31 | Theory | [43] |
| 200 | RS-RbCl | mp-23295 | 0.51 | 1.85 | Theory | [43] |
| 201 | ZB-ZnTe | mp-2176 | 18.86 | 18.92 | Theory | [43] |
| 202 | RS-RbF | mp-11718 | 1.69 | 2.58 | Theory | [43] |
| 203 | RS-RbH | mp-24721 | 2.26 | 4.25 | Theory | [43] |
| 204 | RS-RbI | mp-22903 | 0.06 | 1.75 | Theory | [43] |
| 205 | RS-SrO | mp-2472 | 8.63 | 8.86 | Theory | [43] |

# S3 MatterK,database

## S3.1 The database

A summary of all material entries with thermal conductivity predictions can be found in a separate file.

## S3.2 Thermal conductivity phase diagram for RSS B-C-N chemical system and beyond

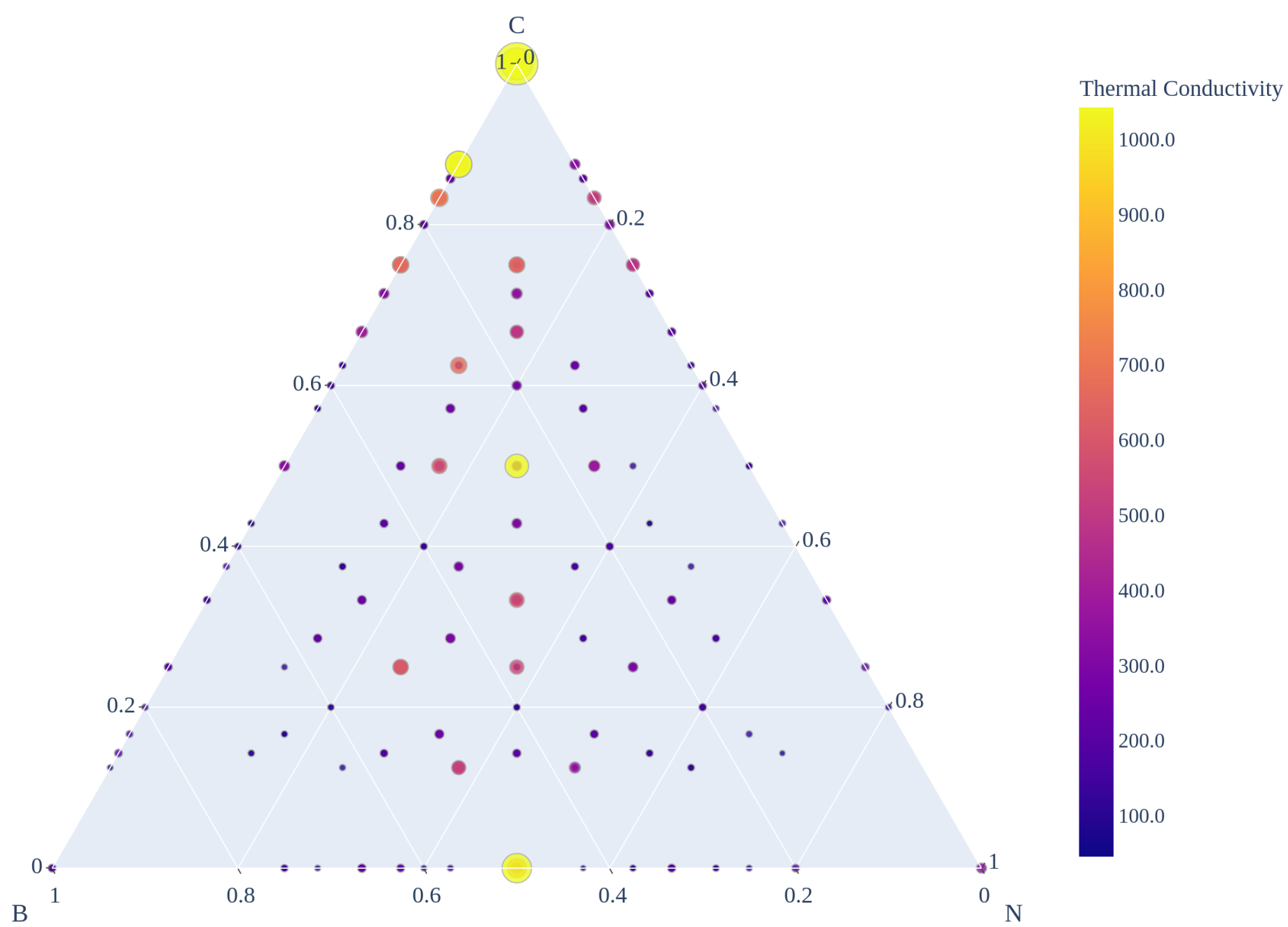

**Fig. S4**: Thermal conductivity phase diagram for B-C-N chemical system with structures generated through random structure search.

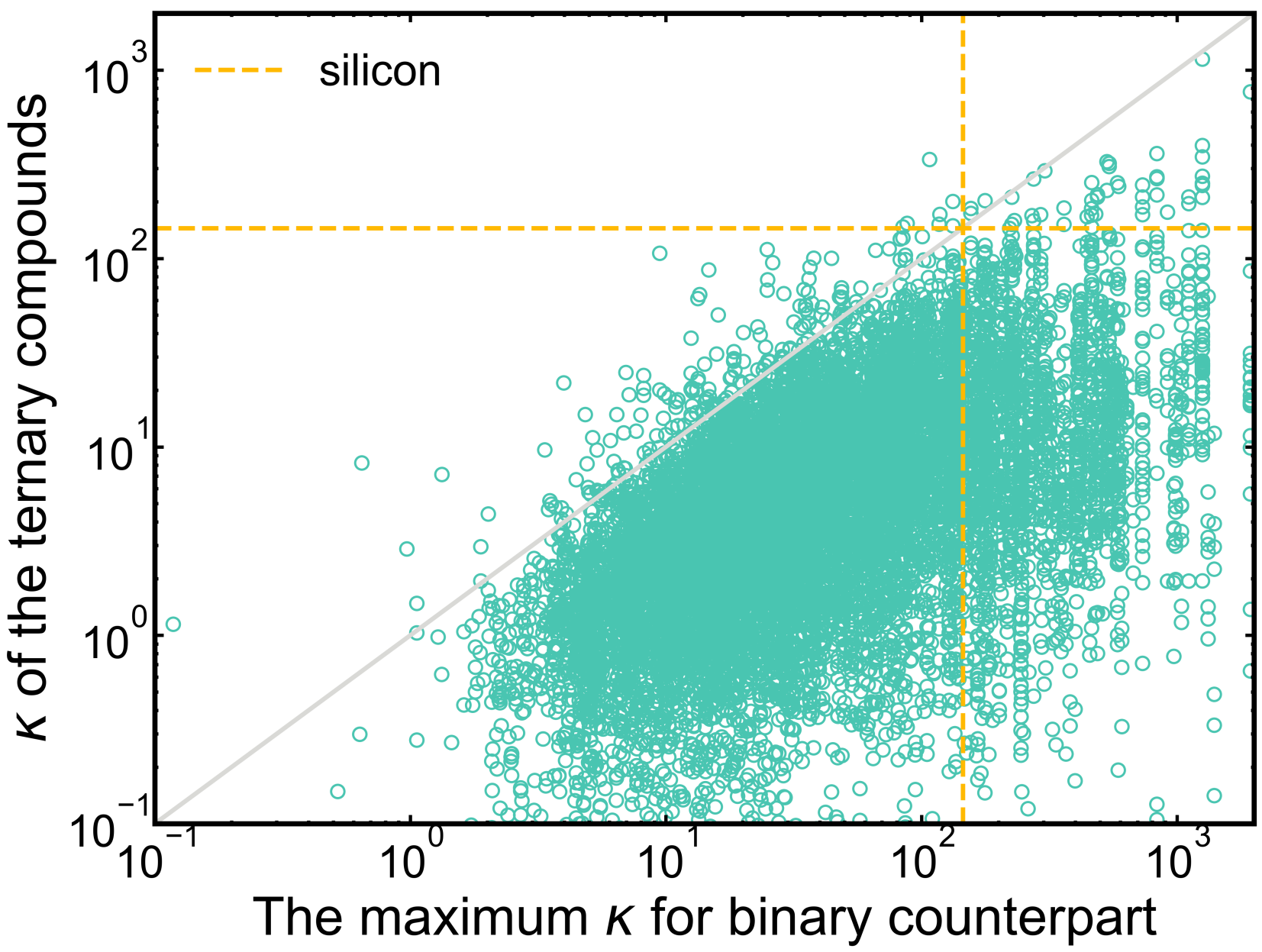

**Fig. S5**: Comparison of the $\kappa$ values of ternary compounds with the maximum $\kappa$ of their corresponding binary counterparts.

### S3.3 Revisit the Slack's rule

Given this extensive data set, we revisit Slack's theory, which postulates that high thermal conductivity in nonmetals correlates with low atomic mass, strong bonding, simple crystal structure, and low anharmonicity [44]. Previous studies have demonstrated that low atomic mass alone is not a sufficient criterion, as materials with large mass ratios, such as BAs and TaN, can also exhibit outstanding thermal transport properties. Instead, the shear modulus-which serves as a proxy for bond strength-emerges as a more robust predictor of high thermal conductivity. The relationship between thermal conductivity and shear modulus (Fig. 3) supports this trend across a diverse range of materials, though with considerable spread. Notably, certain materials in the lower-right region of Fig. 3 exhibit strong elastic constants despite having low thermal conductivity, suggesting potential applications in thermal insulation and mechanical reinforcement.

High symmetry is often considered a key characteristic of materials with high thermal conductivity. However, our analysis of the correlation between thermal conductivity and crystal symmetry operations (Fig. S8) reveals no clear trend, suggesting that symmetry alone is not a decisive factor. For instance, $BC_5$ (mp-1386486), despite having only two symmetry operations, exhibits high thermal conductivity. In this layered material, the interlayer offsets disrupt its symmetry, yet it retains other key characteristics associated with high thermal transport. These findings indicate that, while symmetry can influence thermal conductivity, additional factors-such as bonding strength, phonon dispersion, and scattering mechanisms-play a more dominant role.

## S4 Thermal conductivity distributions

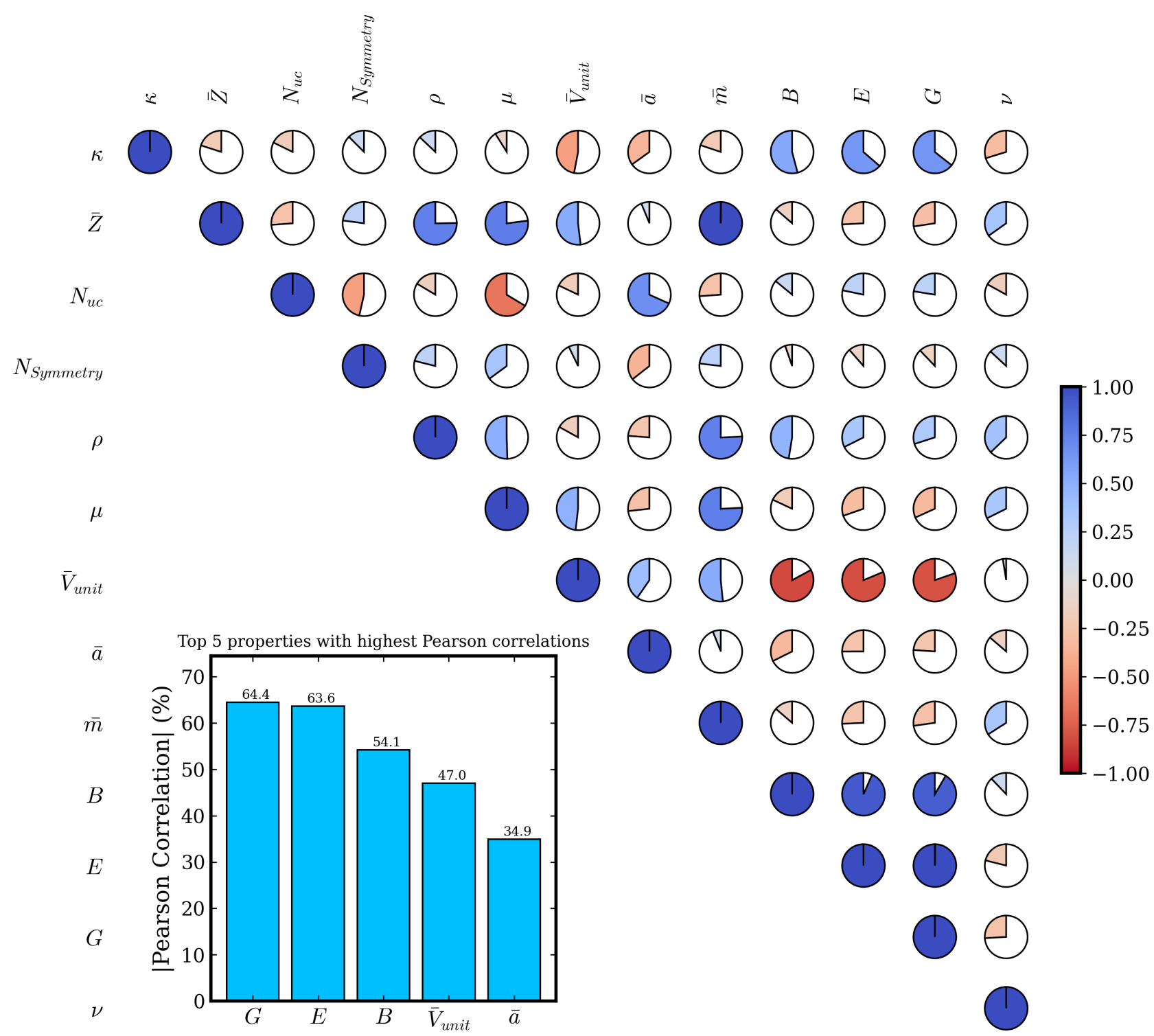

**Fig. S6**: Pearson correlation coefficients between a set of 12 material properties (Table S4) and $\kappa$ computed through MatterSim-BTE for materials in Materials Project database[1]. Each pie chart represents the pairwise correlation coefficient between two properties. The color bar on the right ranges from $-1.0$ (strong negative correlation) to $+1.0$ (strong positive correlation). The bar chart at the lower right highlights the top five properties with the highest Pearson correlations (by absolute magnitude) across all pairs.

**Table S4**: List of material properties used in Pearson correlation analysis. Range indicates the minimum and maximum of the properties calculated with crystal structures from Material Project[1] using MatterSim.

| Descriptions | Symbols | Units | Range |
|---|---|---|---|
| Shear modulus | $G$ | GPa | [0.107, 517] |
| Young's modulus | $E$ | GPa | [0.267, 1122.3] |
| Bulk modulus | $B$ | GPa | [0.178, 463.5] |
| Poisson's ratio | $\nu$ | N/A | [0.003, 0.482] |
| Average lattice constants | $\bar{a}$ | Å | [2.438, 20.716] |
| Average atomic number | $\bar{Z}$ | N/A | [1.0, 94.0] |
| Average atomic masses | $\bar{m}$ | a.m.u | [1.008, 244.1] |
| Normalized volume of the unit-cell | $\bar{V}_{unit}$ | $Å^3$ | [5.031, 583.2] |
| Number of cell symmetry operations | $N_{symmetery}$ | N/A | [1, 48] |
| Number of atoms in the unit-cell | $N_{uc}$ | N/A | [1, 15] |
| Global reduced masses [45] | $\mu$ | a.m.u | [0.084, 244.1] |
| Mass density | $\rho$ | $g\ cm^{-3}$ | [0.020, 22.547] |

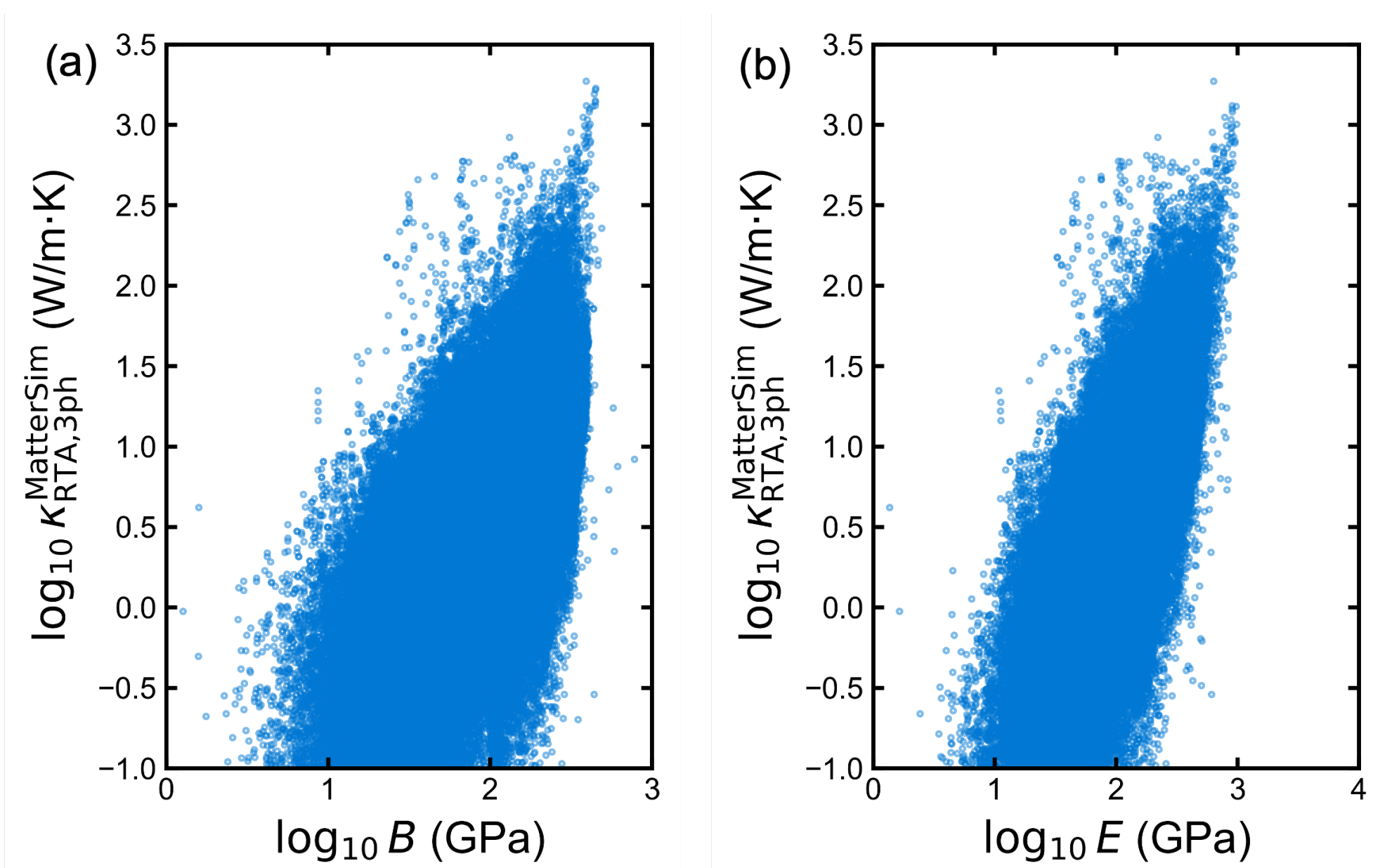

**Fig. S7**: The lattice thermal conductivity (under three-phonon scattering) distribution with respect to the Bulk modulus (B) and Young's modulus (E) for MatterK dataset.

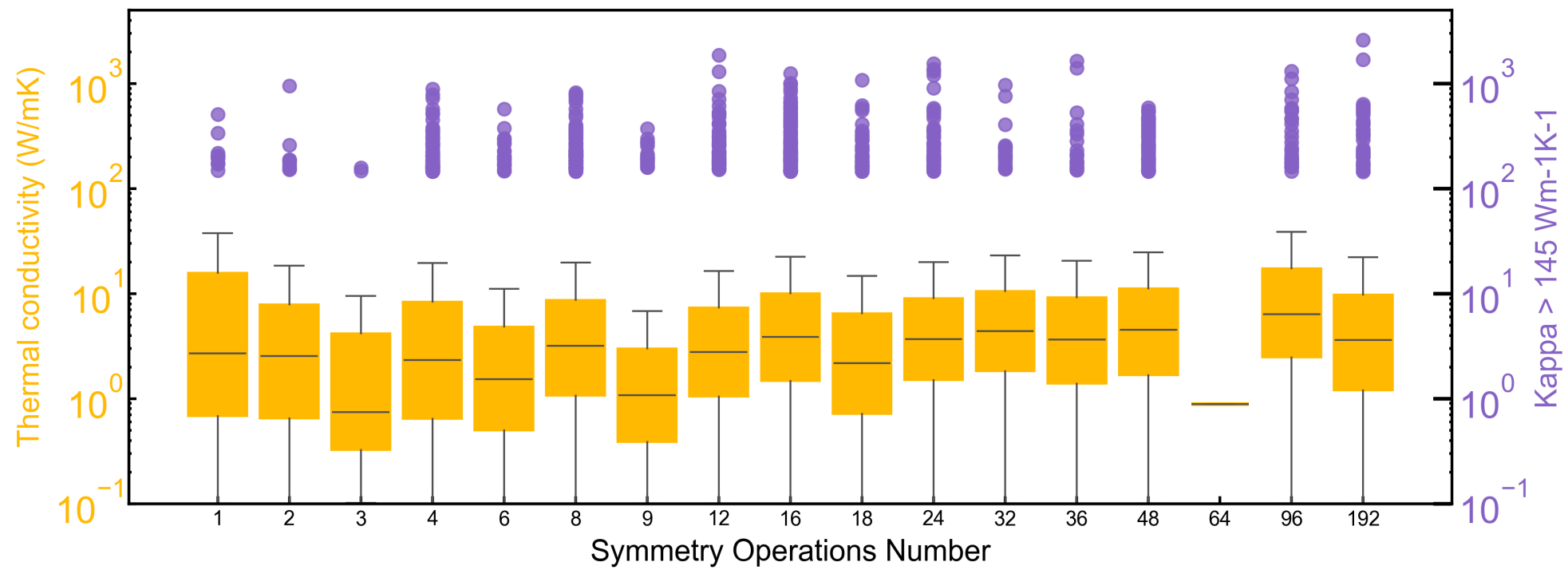

**Fig. S8**: The lattice thermal conductivity (under three-phonon scattering) distribution with respect to symmetry operation number determined through get_symmetry_operations function from pymatgen package[46].

## S5 High thermal conductors

In this section, we listed all high thermal conductor candidates identified in this work. This includes: (1) Previously known high thermal conductors, as listed in Table S5; (2) New high thermal conductors verified by DFT, listed in Table 1 and Table S6 - Table S7. Here, a "new" high thermal conductor is defined as a novel phase of a given chemical composition exhibiting a thermal conductivity exceeding that of silicon ($145\,\mathrm{W\,m^{-1}\,K^{-1}}$). (3) Unverified high thermal conductors predicted via large-scale near-first-principles calculations, performed using MatterSim-accelerated 3ph-BTE and provided in a seperate supporting material.

**Table S5**: Summary of all known high thermal conductors. "band gap" (unit: eV) is the electronic band gap in the PBE level of theory. $E_{\mathrm{hull}}$ represents the energy above hull with the unit of eV/atom. $\kappa$ shown in the table is in the unit of $\mathrm{W\,m^{-1}\,K^{-1}}$. $\kappa_{\mathrm{ref}}$ stands for the reference $\kappa$ values from the literature. The parentheses following the $\kappa$ values indicate that the material is anisotropic, and the value inside the parentheses represents the $\kappa$ values in other directions.

| Systems | mp-id | space group | band gap | $E_{\mathrm{hull}}$ | $\kappa^{\mathrm{MatterKappa}}_{\mathrm{RTA,3ph}}$ | $\kappa_{\mathrm{ref}}$ | $\kappa_{\mathrm{ref}}$ type | Source |
|---|---|---|---|---|---|---|---|---|
| Diamond | mp-66 | Fd$\bar{3}$m | 4.1 | 0.14 | 1683 | 2000 | Exp. | [44] |
| lonsdaleite(hex-C) | mp-47 | P$6_3$/mmc | 3.3 | 0.16 | 1580,1580,1494 | 1629,1629,1400 | DFT-3ph | [47] |
| bct-$C_4$ | mp-1008395 | I4/mmm | 2.2 | 0.34 | 813,813,1293 | 1058,1058,1661 | DFT-3ph | [47] |
| bct-$C_8$ | mp-570002 | Im$\bar{3}$m | 3.0 | 0.77 | 703 | 715 | DFT-3ph | [47] |
| m-carbon | mp-1080826 | C2/m | 3.5 | 0.30 | 992,670,640 | 494,849,622 | DFT-3ph | [47] |
| Z-carbon | mp-1078845 | Cmmm | 3.3 | 0.27 | 1264,887,867 | 1262 | DFT-3ph | [48] |
| c-BN | mp-1639 | F$\bar{4}$3m | 4.5 | 0.08 | 1111 | 768 | Exp. | [49, 50] |
| h-BN | mp-2653 | P$6_3$mc | 5.2 | 0.09 | 853(838) | 602(573) | DFT-3ph | [21] |
| BP | mp-1479 | F$\bar{4}$3m | 1.24 | 0.00 | 275 | 460 | Exp. | [44, 51] |
| BP | mp-1008559 | P$6_3$mc | 1.1 | 0.01 | 267(228) | 395(307) | DFT-3ph-iso | [21] |
| BAs | mp-10044 | F$\bar{4}$3m | 1.2 | 0.08 | 849 | 1200 | Exp. | [52–54] |
| BAs | mp-984718 | P$6_3$mc | 1.1 | 0.09 | 646,646,550 | 1440,1440,1090 | DFT-3ph-RTA | [21] |
| $B_2$AsP | mp-1008528 | P$\bar{4}$m2 | 1.1 | 0.07 | 268,268,185 | 372,372,308 | DFT-3ph | [55] |
| c-SiC | mp-8062 | F$\bar{4}$3m | 1.4 | 0.00 | 465 | 490 | Exp. | [56] |
| 2H-SiC | mp-7140 | P$6_3$mc | 2.3 | 0.004 | 423(389) | 497 | DFT-3ph-iso | [57] |
| 4H-SiC | mp-11714 | P$6_3$mc | 2.2 | 0.00 | 330(221) | 415(345) | Exp. | [58] |
| 6H-SiC | mp-7631 | P$6_3$mc | 2.0 | 0.00 | 296(256) | 390(320) | Exp. | [58] |
| BeO | mp-1778 | F$\bar{4}$3m | 6.9 | 0.007 | 352 | 358 | DFT-3ph-RTA | [21] |
| BeO | mp-2542 | P$6_3$mc | 7.4 | 0.00 | 278(288) | 370 | Exp. | [59] |
| AlN | mp-1700 | F$\bar{4}$3m | 3.3 | 0.02 | 233 | 212 | DFT-3ph-RTA | [21] |
| AlN | mp-661 | P$6_3$mc | 4.05 | 0.00 | 221(235) | 320 | Exp. | [44] |
| GaN | mp-830 | F$\bar{4}$3m | 1.6 | 0.007 | 199 | 181 | DFT-3ph | [21] |
| GaN | mp-804 | P$6_3$mc | 1.7 | 0.00 | 204.5(209.0) | 200 | Exp. | [58] |
| $\theta$-TaN | mp-1459 | P$\bar{6}$m2 | 0.0 | 0.02 | 562(1057) | 995(820) | DFT-4ph-epc-el | [60] |
| $BC_5$ | mp-1018649 | P3m1 | 0.0 | 0.28 | 152(171) | 165(169) | DFT-3ph | [61] |
| Si | mp-149 | Fd$\bar{3}$m | 0.6 | 0.00 | 145 | 142 | Exp. | [62] |
| WC | mp-1894 | P$\bar{6}$m2 | 0.0 | 0.00 | 1963.6(1733.2) | 204(249) | DFT-3ph-epc-el | [63] |
| BiB | mp-1006880 | F$\bar{4}$3m | 0.1 | 0.82 | 587 | 347 | DFT-3ph | [64] |
| GeC | mp-1002164 | F$\bar{4}$3m | 1.7 | 0.44 | 548 | 1517 | DFT-3ph | [20] |
| 2H-GeC | mp-1184550 | P$6_3$mc | 2.4 | 0.44 | 536,536,498 | 1350,1350,1050 | DFT-3ph | [65] |
| $BC_2N$ | mp-629458 | Pmm2 | 1.7 | 0.54 | 677(449) | 395 | DFT-3ph | [66] |
| MoC | mp-2305 | P$\bar{6}$m2 | 0.0 | 0.00 | 573(494) | 110 | DFT-3ph-epc-el | [63] |
| BSb | mp-997618 | F$\bar{4}$3m | 0.8 | 0.27 | 469 | 465 | DFT-3ph-iso | [3] |
| $BC_2N$ | mp-30148 | P$222_1$ | 2.1 | 0.55 | 491(439,424) | 461 | DFT-3ph | [66] |
| BeTe | mp-252 | F$\bar{4}$3m | 2.0 | 0.00 | 174 | 286 | DFT-3ph-RTA | [21] |
| BeSe | mp-1541 | F$\bar{4}$3m | 2.7 | 0.00 | 168 | 633 | DFT-3ph | [20] |
| Be | mp-87 | P$6_3$/mmc | 0.0 | 0.00 | 181(158) | 200 | Exp. | [67] |

**Table S6**: Potential high-$\kappa$ candidates verified by DFT at the level of BTE with three-phonon scattering. The structures are sourced from the Materials Project database (labeled with MP IDs) and an in-house random structure search database (labeled with RSS). The upper panel includes structures with a band gap, while the lower panel consists of structures without a band gap. The energy above the convex hull, $E_{\mathrm{hull}}$, is reported in eV/atom, where higher values indicate greater thermodynamic instability for synthesis. Thermal conductivity values, $\kappa$, are given in $\mathrm{W\,m^{-1}\,K^{-1}}$. $\kappa^{\mathrm{MatterSim}}_{\mathrm{RTA,3ph}}$ represents values computed using MatterSim under the relaxation time approximation (RTA), considering only three-phonon scattering. $\kappa^{\mathrm{DFT}}_{\mathrm{LBTE}}$ denotes reference values obtained in this work via DFT using the full solution of the linearized phonon Boltzmann equation (LBTE). PBE functional is adopted in the DFT calculations unless otherwise specified. The Type column specifies the scattering processes considered in the DFT-BTE calculations: "3ph" (three-phonon scattering) contribution to thermal conductivity.

| Systems | ID | Space Group | $E_{\mathrm{hull}}$ | $\kappa^{\mathrm{MatterSim}}_{\mathrm{RTA,3ph}}$ xx\|yy\|zz | $\kappa^{\mathrm{DFT}}_{\mathrm{LBTE}}$ xx\|yy\|zz | Type |
|---|---|---|---|---|---|---|
| $B_6O$ | mp-1346 | R-3m | 0.00 | 196\|216\|**221** | **279**\|279\|206 | 3ph |
| $B_6P$ | mp-28395 | R-3m | 0.00 | 153\|157\|**159** | **209**\|209\|189 | 3ph |
| $AlGaN_2$ | mp-1228894 | P3m1 | 0.01 | **159**\|159\|126 | **181**\|181\|144 | 3ph |
| SiGe | mp-1219182 | F-43m | 0.02 | 190\|190\|190 | 150\|150\|150 | 3ph |
| AlSiCN | mp-1227998 | P3m1 | 0.04 | **228**\|228\|145 | **317**\|317\|275 | 3ph |
| $WN_2$ | mp-999549 | P-6m2 | 0.09 | 219\|219\|**241** | 436\|436\|**517** | 3ph |
| $WN_2$ | mp-1077232 | $P6_3$/mmc | 0.09 | 181\|181\|**227** | 300\|300\|**303** | 3ph |
| BP | RSS | R-3m | 0.11 | 203\|211\|**215** | **426**\|426\|342 | 3ph |
| BP | RSS | P-6m2 | 0.14 | **226**\|226\|189 | **388**\|388\|244 | 3ph |
| $BeSiN_2$ | mp-1227309 | P3m1 | 0.17 | **211**\|211\|80 | **280**\|280\|102 | 3ph |
| BN | mp-13151 | $P4_2$/mnm | 0.18 | **648** \| 348 \| 348 | 451\|451\|**1023** | 3ph |
| $PtN_2$ | mp-1095618 | Pa-3 | 0.21 | 144\|144\|144 | 216\|216\|216 | 3ph |
| BN | mp-644751 | Pnma | 0.27 | **313**\|254\|228 | 266\|**500**\|356 | 3ph |
| BN | mp-1077506 | Imm2 | 0.30 | 141\|**251**\|249 | 184\|304\|**324** | 3ph |
| CH† | mp-1079612 | I2_13 | 0.32 | 185\|185\|185 | 536\|536\|536 | 3ph |
| $CN_2$ | mp-1018655 | P-3m1 | 0.6 | **362**\|362\|14 | **427**\|427\|11 | 3ph |
| $BC_2N$ | mp-1078541 | C2/m | 0.64 | **420**\|104\|303 | 400\|**891**\|632 | 3ph |
| $CN_2$ | mp-1009818 | I-4m2 | 0.73 | 223\|228\|**231** | **399**\|399\|216 | 3ph |
| $BC_2N$ | mp-1008523 | P-4m2 | 1.00 | 194\|194\|194 | **552**\|552\|487 | 3ph |
| $Be_2CoNi$ | mp-867271 | Fm-3m | 0.00 | 292\|292\|292 | 519\|519\|519 | 3ph |
| NbB | mp-2580 | Cmcm | 0.00 | 115\|**195**\|145 | **400**\|319\|312 | 3ph |
| $MnV_2Cr$ | mp-864953 | Fm-3m | 0.00 | 169\|169\|169 | 392\|392\|392 | 3ph |
| HfS | mp-1206743 | P-6m2 | 0.00 | 82\|82\|**302** | 247\|247\|**352** | 3ph |
| VCr | RSS | Pm-3m | 0.00 | 254\|254\|254 | 332\|332\|332 | 3ph |
| TaP | mp-1067587 | $I4_1$md | 0.00 | **152**\|152\|59 | **319**\|319\|136 | 3ph |
| MnV | mp-316 | Pm-3m | 0.00 | 149\|149\|149 | 706\|706\|706 | 3ph |
| VB | RSS | Fm-3m | 0.00 | 504\|504\|504 | 228\|228\|228 | 3ph |
| $Al_2IrOs$ | mp-866284 | Fm-3m | 0.00 | 335\|335\|335 | 192\|192\|192 | 3ph |
| TaRe | RSS | Pm-3m | 0.00 | 146\|146\|146 | 191\|191\|191 | 3ph |
| VN | mp-1018027 | P-6m2 | 0.00 | 152\|152\|**208** | 157\|**190**\|157 | 3ph |
| TaW | RSS | Pm-3m | 0.00 | 146\|146\|146 | 190\|190\|190 | 3ph |
| $B_2W$ | RSS | R-3m | 0.00 | **179**\|168\|164 | 118\|118\|158 | 3ph |
| $MoWC_2$ | mp-1221393 | Pmm2 | 0.00 | **766**\|568\|201 | **921**\|549\|207 | 3ph |
| ReN | RSS | R3m | 0.01 | 158\|158\|**242** | 181\|181\|146 | 3ph |
| $HoTmAl_2$ | mp-1184827 | Fm-3m | 0.02 | 241\|241\|241 | 247\|247\|247 | 3ph |
| $DyErAl_2$ | mp-1183795 | Fm-3m | 0.02 | 230\|230\|230 | 207\|207\|207 | 3ph |

| Continuation of Table S6 | | | | | | |
|---|---|---|---|---|---|---|
| Systems | ID | Space Group | $E_{\text{hull}}$ | $\kappa^{\text{MatterSim}}_{\text{RTA,3ph}}$ xx\|yy\|zz | $\kappa^{\text{DFT}}_{\text{LBTE}}$ xx\|yy\|zz | Type |
| CrC | mp-1018050 | P-6m2 | 0.08 | 140\|140\|**154** | 174\|174\|**224** | 3ph |
| $B_2CN$ | mp-1079333 | Pmma | 0.24 | 402\|**525**\|227 | 373\|**432**\|215 | 3ph |
| $OsN_2$ | mp-973935 | P6/mmm | 0.25 | 98\|98\|**1322** | 97\|97\|**931** | 3ph |
| $BC_5$ | mp-1077125 | I-4m2 | 0.25 | **535**\|535\|480 | **393**\|393\|242 | 3ph |
| $BC_7$ | mp-1079046 | Pmm2 | 0.25 | **1003**\|699\|955 | **231**\|206\|163 | 3ph |
| ReC | mp-1009735 | P-6m2 | 0.27 | **369**\|369\|328 | 225\|225\|**362** | 3ph |
| $BC_7$ | mp-1095030 | P-43m | 0.28 | 1201\|1201\|1201 | 481\|481\|481 | 3ph |
| TiN | mp-998908 | F-43m | 0.30 | 312\|312\|312 | 199\|199\|199 | 3ph |
| C | mp-1008374 | Cmmm | 0.44 | **564**\|114\|368 | 237\|620\|**1157** | 3ph |
| BiB | mp-1183440 | $P6_3mc$ | 0.84 | 289\|289\|**355** | **168**\|168\|138 | 3ph |
| End of Table S6 | | | | | | |

∗ LDA functional is adopted in DFT calculations.

† $K_4$ phase of the carbon-hydrogen compound identified in Ref. 68.

**Table S7**: Potential high-$\kappa$ candidates verified by DFT at the level of BTE with multiple scattering channels considered. The structures are sourced from the Materials Project database (labeled with MP IDs) and an in-house random structure search database (labeled with RSS). The energy above the convex hull, $E_{\text{hull}}$, is reported in eV/atom, where higher values indicate greater thermodynamic instability for synthesis. Thermal conductivity values, $\kappa$, are given in $\mathrm{W\,m^{-1}\,K^{-1}}$. $\kappa^{\text{MatterSim}}_{\text{RTA,3ph}}$ represents values computed using MatterSim under the relaxation time approximation (RTA), considering only three-phonon scattering. $\kappa^{\text{DFT}}_{\text{LBTE}}$ denotes reference values obtained in this work via DFT using the full solution of the linearized phonon Boltzmann equation (LBTE). PBE functional is adopted in the DFT calculations unless otherwise specified. The Type column specifies the scattering processes considered in the DFT-BTE calculations: "3ph" (three-phonon scattering), "4ph" (four-phonon scattering), "iso" (phonon-isotope scattering), "eph" (electron-phonon coupling), and "el" (electronic) contribution to thermal conductivity.

| Systems | ID | Space Group | $E_{\text{hull}}$ | $\kappa^{\text{MatterSim}}_{\text{RTA,3ph}}$ xx\|yy\|zz | $\kappa^{\text{DFT}}_{\text{LBTE}}$ xx\|yy\|zz | Type |
|---|---|---|---|---|---|---|
| TaP* | mp-1187244 | P-6m2 | 0.003 | 153\|153\|**505** | 230\|230\|**366** | 3ph+4ph+iso+eph+el |
| MnV | mp-316 | Pm-3m | 0.00 | - | 205\|205\|205 | 3ph+4ph+iso+eph+el |
| MnV* | mp-316 | Pm-3m | 0.00 | - | 243\|243\|243 | 3ph+4ph+iso+eph+el |
| ReB | RSS | P-6m2 | 0.00 | **231**\|231\|199 | 90\|90\|**158** | 3ph+iso+eph+el |
| $MoWC_2$ | mp-1221393 | Pmm2 | 0.00 | - | **154**\|117\|91 | 3ph+iso+eph+el |
| NbN | mp-2634 | P-6m2 | 0.00 | 365\|365\|**511** | 122\|122\|**125** | 3ph+iso+eph+el |
| ZrN | mp-1352 | Fm-3m | 0.00 | 147\|147\|147 | 69\|69\|69 | 3ph+iso+eph+el |
| BeSe | mp-1541 | F-43m | 0.00 | 163\|163\|163 | 58\|58\|58 | 3ph+4ph+iso |
| CuBe | mp-2323 | Pm-3m | 0.00 | 404\|404\|404 | 44\|44\|44 | 3ph+4ph+iso |
| HfP | RSS | Fm-3m | 0.02 | 635\|635\|635 | 42\|42\|42 | 3ph+4ph+iso |
| TaN* | mp-1459 | P-6m2 | 0.03 | 540\|540\|**1040** | **1023**\|1023\|806 | 3ph+4ph+iso+eph+el |
| ZrP | mp-930 | Fm-3m | 0.04 | 605\|605\|605 | 31\|31\|31 | 3ph+iso+eph+el |
| WC | RSS | $I4_1md$ | 0.14 | **833**\|833\|525 | **294**\|294\|198 | 3ph+iso+eph+el |
| TaN | mp-570604 | $P6_3/mmc$ | 0.24 | 159\|159\|**186** | 48\|48\|**54** | 3ph+iso+eph+el |
| $OsN_2$ | mp-973935 | P6/mmm | 0.25 | - | 112\|112\|**435** | 3ph+4ph+iso+eph+el |

| Continuation of Table S7 | | | | | | |
|---|---|---|---|---|---|---|
| Systems | ID | Space Group | $E_{\text{hull}}$ | $\kappa_{\text{RTA,3ph}}^{\text{MatterSim}}$ | $\kappa_{\text{LBTE}}^{\text{DFT}}$ | Type |
| | | | | xx\|yy\|zz | xx\|yy\|zz | |
| $BC_7$ | mp-1095030 | P-43m | 0.28 | - | 74\|74\|74 | 3ph+iso+eph+el |
| NbB | RSS | Fm-3m | 0.29 | 2588\|2588\|2588 | 104\|104\|104 | 3ph+iso+eph+el |
| BMo | RSS | Fm-3m | 0.36 | 417\|417\|417 | 40\|40\|40 | 3ph+iso+eph+el |
| ZrB | mp-451 | Fm-3m | 0.37 | 533\|533\|533 | 74\|74\|74 | 3ph+iso+eph+el |
| BiB | mp-1006880 | F-43m | 0.82 | 585\|585\|585 | 168\|168\|168 | 3ph+4ph+iso |
| End of Table S7 | | | | | | |

$*$ LDA functional is adopted in DFT calculations.

† $K_4$ phase of the carbon-hydrogen compound identified in Ref. 68.

# S6 Metals with novel thermal transport mechanism

## S6.1 Detailed analysis for MnV, TiFe & ScCo

As discussed in the main text, the phonon dispersion relations reveal that all three acoustic phonon branches of MnV are degenerate at the high-symmetry $R$ point. Additionally, the energy difference between the acoustic and optical phonon branches at the $R$ point is small, leading to extended phonon lifetimes for frequencies beyond 8 THz. These prolonged phonon lifetimes result in a $\kappa_{ph}$ as high as $661\,\mathrm{W\,m^{-1}\,K^{-1}}$, more than four times that of bulk Si, with isotopic scattering included in the calculations [38, 62, 69].

To further investigate the origin of this elevated $\kappa_{ph}$ within the Boltzmann transport equation (BTE) framework considering three-phonon scattering, we analyzed the norm of the group velocity, phonon lifetimes, and modal conductivity ($\kappa_{\text{per mode}}$) along high-symmetry directions in the Brillouin zone (Fig. S12). The collective analysis of these three quantities indicates that, in addition to the dominant contributions near the $\Gamma$ point, phonon modes around the $R$ point significantly enhance $\kappa_{\text{ph}}$. A similar trend has previously been reported in *bcc* elementary metals such as W, Mo, and Cr [70].

Further calculations account for four-phonon and electron-phonon scattering processes. Electron-phonon coupling reduces $\kappa_{\text{ph}}$ to $153\,\mathrm{W\,m^{-1}\,K^{-1}}$, while four-phonon scattering further lowers it to $119\,\mathrm{W\,m^{-1}\,K^{-1}}$. A detailed analysis of MnV is presented in the main text.

In addition to MnV, similar analyses have been conducted for the binary intermetallic materials TiFe and ScCo, which also crystallize in the *bcc* structure. Notably, TiFe emerges as a promising candidate, with $\kappa_{\text{ph}} \simeq 200$ when only three-phonon (3ph) scattering is considered. Like MnV, TiFe exhibits phonon degeneracy at the $R$ point in the Brillouin zone, though the energy difference between its optical and acoustic phonons is larger. Consequently, the scattering rate is enhanced for phonons near $R$ point, and the phonon lifetime is suppressed due to the increased phase space available for energy- and momentum-conserving three-phonon processes.

**Table S8**: Summary of space group symmetry, lattice constants as well as $\kappa_{ph}$ for metallic compounds along each axis. The unit of $\kappa_{ph}$ is in $\mathrm{W\,m^{-1}\,K^{-1}}$.

| materials | space group | a | b | c | $\kappa_{xx}$ | $\kappa_{yy}$ | $\kappa_{zz}$ | $\kappa_{type}$ |
|---|---|---|---|---|---|---|---|---|
| MnV | Pm-3m | 2.875 Å | 2.875 Å | 2.875 Å | 119 | 119 | 119 | 3ph + 4ph + eph + iso |
| TiFe | Pm-3m | 2.956 Å | 2.956 Å | 2.956 Å | 263 | 263 | 263 | 3ph |
| ScCo | Pm-3m | 3.120 Å | 3.120 Å | 3.120 Å | 38.1 | 38.1 | 38.1 | 3ph |
| $OsN_2$ | P6/mmm | 2.836 Å | 2.836 Å | 4.956 Å | 24.0 | 24.0 | 212 | 3ph + 4ph + eph |
| TaP | P$\bar{6}$m2 | 3.266 Å | 3.266 Å | 3.302 Å | 201 | 201 | 316 | 3ph + 4ph + eph |

This trend is further confirmed by the analysis of ScCo, which distinguishes itself from MnV and TiFe by exhibiting an even larger energy difference between its optical and acoustic phonons near $R$ pint, leading to a further reduction in $\kappa_{\mathrm{ph}}$.

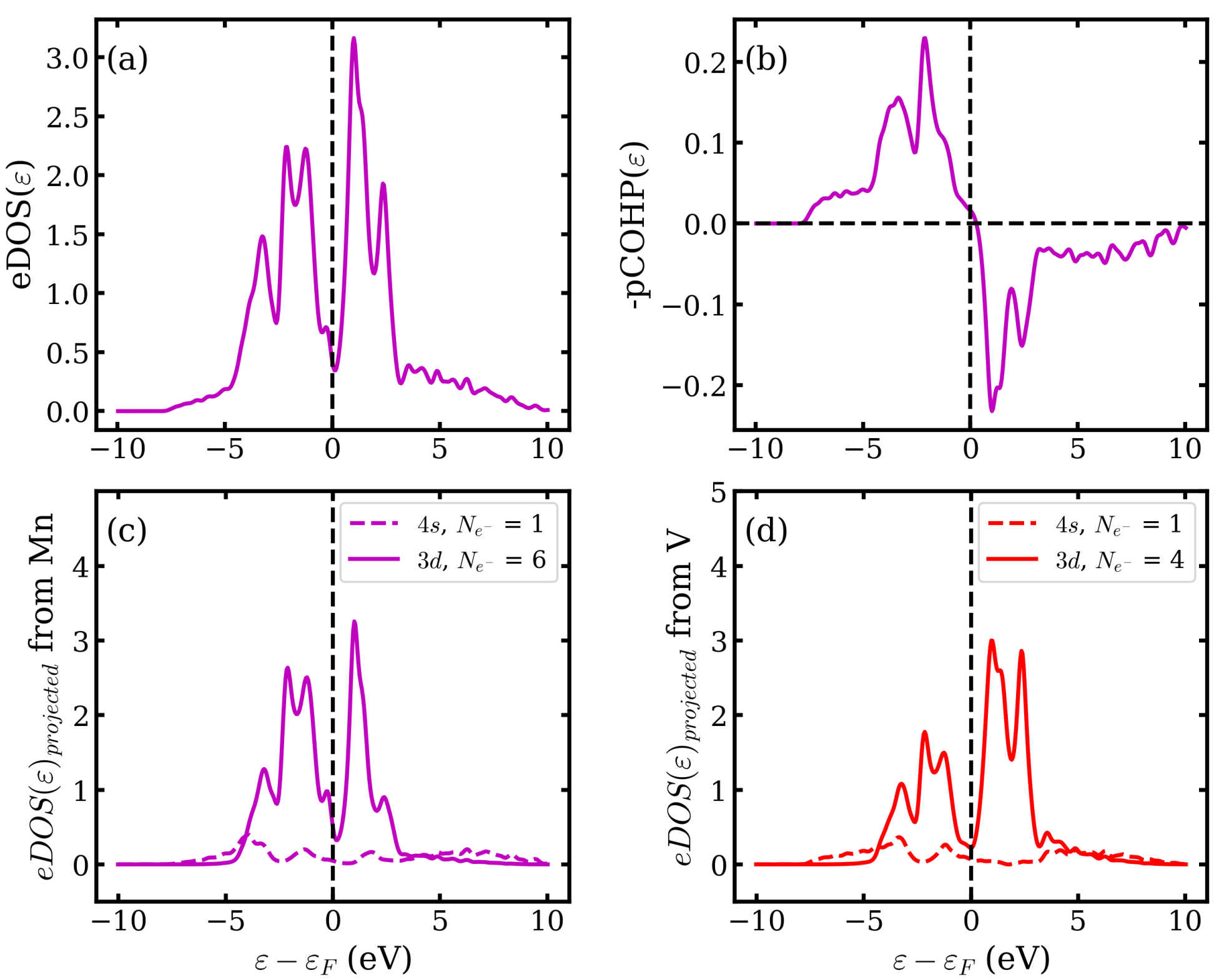

**Fig. S9**: Total electronic density of state (eDOS) (a) and projected crystal orbital Hamilton population (pCOHP) (b) for MnV in the Pm-3m phase. (c) and (d) represent the projection of electronic DOS ($eDOS(\varepsilon)_{projected}$) for Mn and V atoms. $N_{e^-}$ symbolizes number of electrons contributing to bonding in $4s$ and $3d$ orbitals, obtained by integrating the $eDOS(\varepsilon)_{projected}$ up to Fermi energy ($\varepsilon_F$).

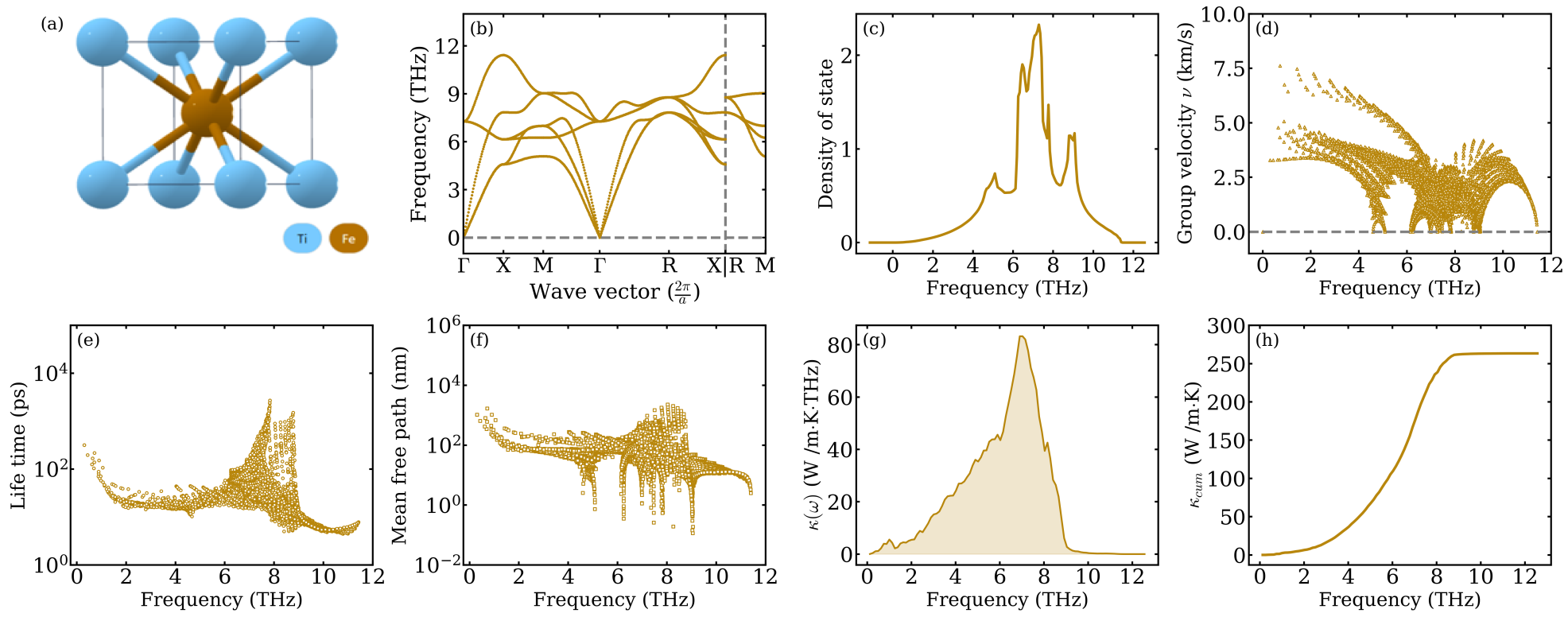

**Fig. S10**: Unit-cell structure of TiFe in Pm-3m phase with its phononic and thermal transport properties. When accounting for three-phonon scattering, $\kappa_{cum}$ reaches to 263 W m$^{-1}$ K$^{-1}$.

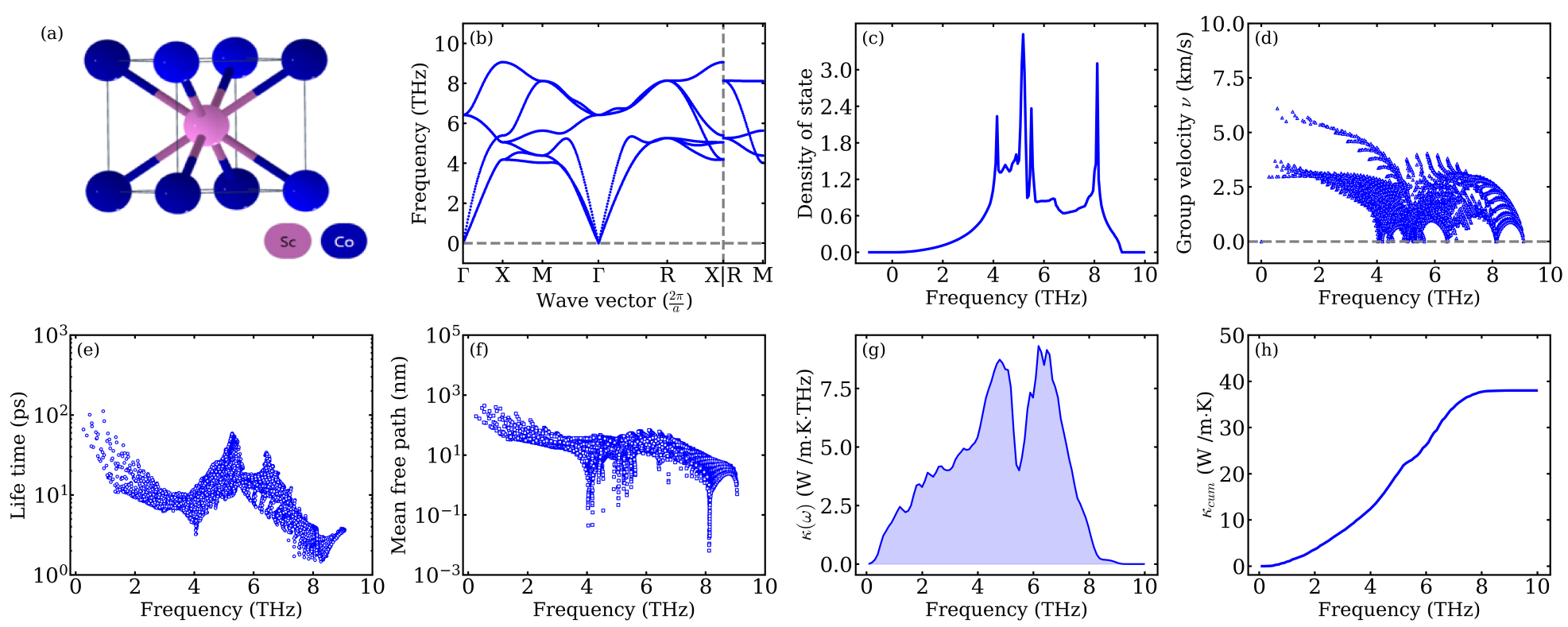

**Fig. S11**: Unit-cell structure of ScCo in Pm-3m phase with phononic and its thermal transport properties. Upon including three-phonon scattering, $\kappa_{cum}$ reaches to 38.1 W m$^{-1}$ K$^{-1}$

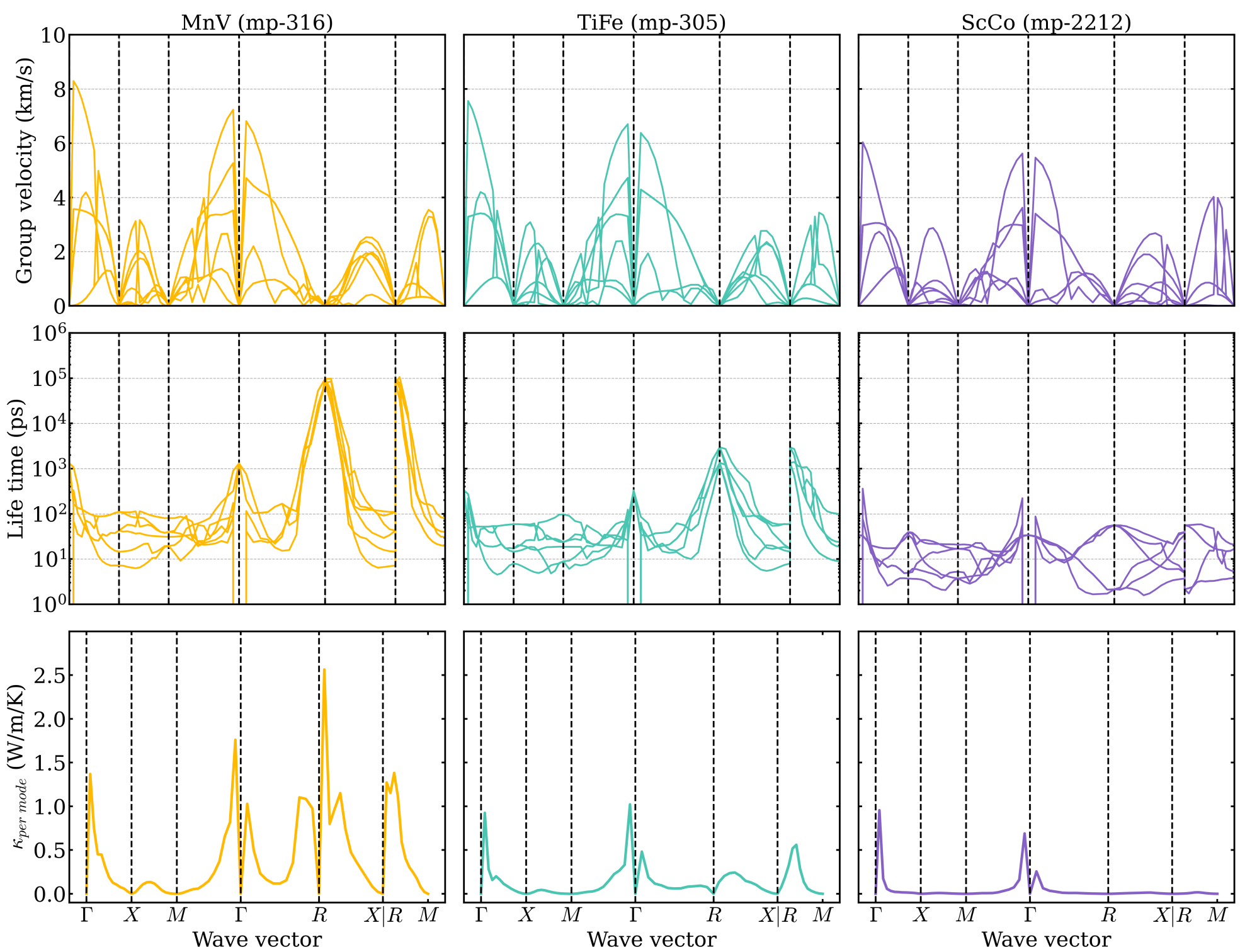

**Fig. S12**: Norm of phonon group velocities, life time and per mode thermal conductivity ($\kappa_{per\ mode}$) for MnV, TiFe and ScCo. Lattice dynamic calculations are performed with DFT-based interatomic force constants using a $4 \times 4 \times 4$ super-cell and a $q$-grid of $32 \times 32 \times 32$. Note that the phonon life time is computed with tetrahedral method and used in tandem with RTA method to obtain $\kappa_{per\ mode}$, as implemented in phono3py[71].